\documentclass[numberedappendix]{emulateapj}
\usepackage{natbib}
\usepackage{graphicx}
\usepackage{color}
\usepackage{bm}

\usepackage{enumerate}

\begin{document}
\shorttitle{Simple foreground cleaning algorithm for detecting $B$-mode
polarization of CMB}
\title{Simple foreground cleaning algorithm for detecting primordial 
$B$-mode polarization of the cosmic microwave background}
\author{
{{Nobuhiko Katayama}}\altaffilmark{1} and {{Eiichiro Komatsu}}\altaffilmark{2}}
\altaffiltext{1}{{Institute of Particle and Nuclear Studies, %
                    High Energy Accelerator Research Organization and
		    School of High Energy Accelerator Science, The
		    Graduate University for Advanced Studies (Sokendai),
		    Tsukuba, Ibaraki, 305-0801, Japan},
		    nobu.katayama@kek.jp} 
\altaffiltext{2}{{Texas Cosmology Center and Department of Astronomy, Univ. of Texas, Austin, Dept. of Astronomy, %
                    2511 Speedway, RLM 15.306, Austin, TX 78712}}
\begin{abstract}
 We reconsider the pixel-based, ``template'' polarized foreground removal method
 within the context of a next-generation,
 low-noise, low-resolution (0.5 degree FWHM) space-borne experiment
 measuring the cosmological $B$-mode polarization signal in the cosmic
 microwave background (CMB). This method was
first applied to polarized data
by 
 the {\sl Wilkinson Microwave Anisotropy Probe} ({\sl WMAP}) team and
 further studied by Efstathiou et al. We need at least 3 frequency
 channels: one is used for extracting the CMB signal, whereas the other
 two are used to estimate the spatial distribution of the polarized dust and
 synchrotron emission. No extra data from non-CMB experiments or models are used.
 We extract the tensor-to-scalar ratio ($r$) from
 simulated sky maps
 outside the standard polarization mask (P06) of {\sl WMAP}
 consisting of CMB, noise ($2~\mu$K~arcmin), and a
 foreground model, and find that, even for the 
 simplest 3-frequency configuration with 60, 100, and 240~GHz, the
 residual bias in $r$ is as small as $\Delta r\approx 0.002$. This bias
 is dominated 
 by the residual synchrotron emission due to spatial
 variations of the synchrotron spectral index.
With an extended mask with $f_{sky}=0.5$, the bias 
is reduced further down to $<0.001$. 
\end{abstract}
\section{Introduction}
\label{sec:introduction}
Why study the $B$-mode polarization of the cosmic microwave background (CMB)?
Detection of the primordial gravitational waves generated during
inflation would give us a direct insight into the physical condition of
the universe when the energy scale was close to the grand unification scale,
$\sim 10^{16}$~GeV \citep[see][for a recent review and references
therein]{liddle/lyth:PDP}. 
While a direct detection of the primordial
gravitational waves using, e.g., laser interferometers, seems not
possible with the present-day technology, an {\it indirect} detection using
the  $B$-mode polarization of the CMB
\citep{seljak/zaldarriaga:1997,kamionkowski/kosowsky/stebbins:1997} 
may be possible in the near future (most optimistically, within a
few years), provided that the energy scale of
inflation at which the observed gravitational waves were generated
was indeed as high as the grand unification scale.

We often characterize the amplitude of gravitational waves (also known
as tensor perturbations) using the so-called ``tensor-to-scalar ratio,''
which is conventionally defined as
\begin{equation}
 r\equiv \frac{2\langle |h^{+}_{{\mathbf k}}|^2+|h^{\times}_{{\mathbf k}}|^2\rangle}{\langle |{\cal R}_{\mathbf{k}}|^2\rangle},
\end{equation}
where $h^{+}_{\mathbf k}$ and $h^{\times}_{\mathbf k}$ are the Fourier
transform of the amplitudes of two linear polarization states of
gravitational waves, and ${\cal R}_{\mathbf k}$ is the primordial
curvature perturbation, which is a scalar perturbation (hence the name,
``tensor-to-scalar ratio''). It is ${\cal R}_{\mathbf k}$ that seeded
the observed structure in the universe, as well as the dominant
component of the observed CMB temperature anisotropy \citep[see][for a
recent review and references therein]{weinberg:COS}. 

The dominant, scalar part of the temperature anisotropy generates
radial and tangential polarization patterns around hot and cold
spots \citep{coulson/crittenden/turok:1994}. This is called the $E$-mode
polarization, and has been detected 
with high statistical significance
\citep{brown/etal:2009,chiang/etal:2010,larson/etal:prep,komatsu/etal:prep,quiet:prep}. However, 
the $B$-mode polarization, which 
cannot be generated by the scalar perturbations but can be generated by
the tensor perturbations, has not been found
yet. The current 95\% upper limit on the tensor-to-scalar ratio is
$r<0.24$, which mainly comes from the upper limit on the tensor
contribution to the temperature anisotropy on large angular scales
\citep{komatsu/etal:prep}.  

\begin{figure}[t]
\centering \noindent
\includegraphics[width=8.5cm]{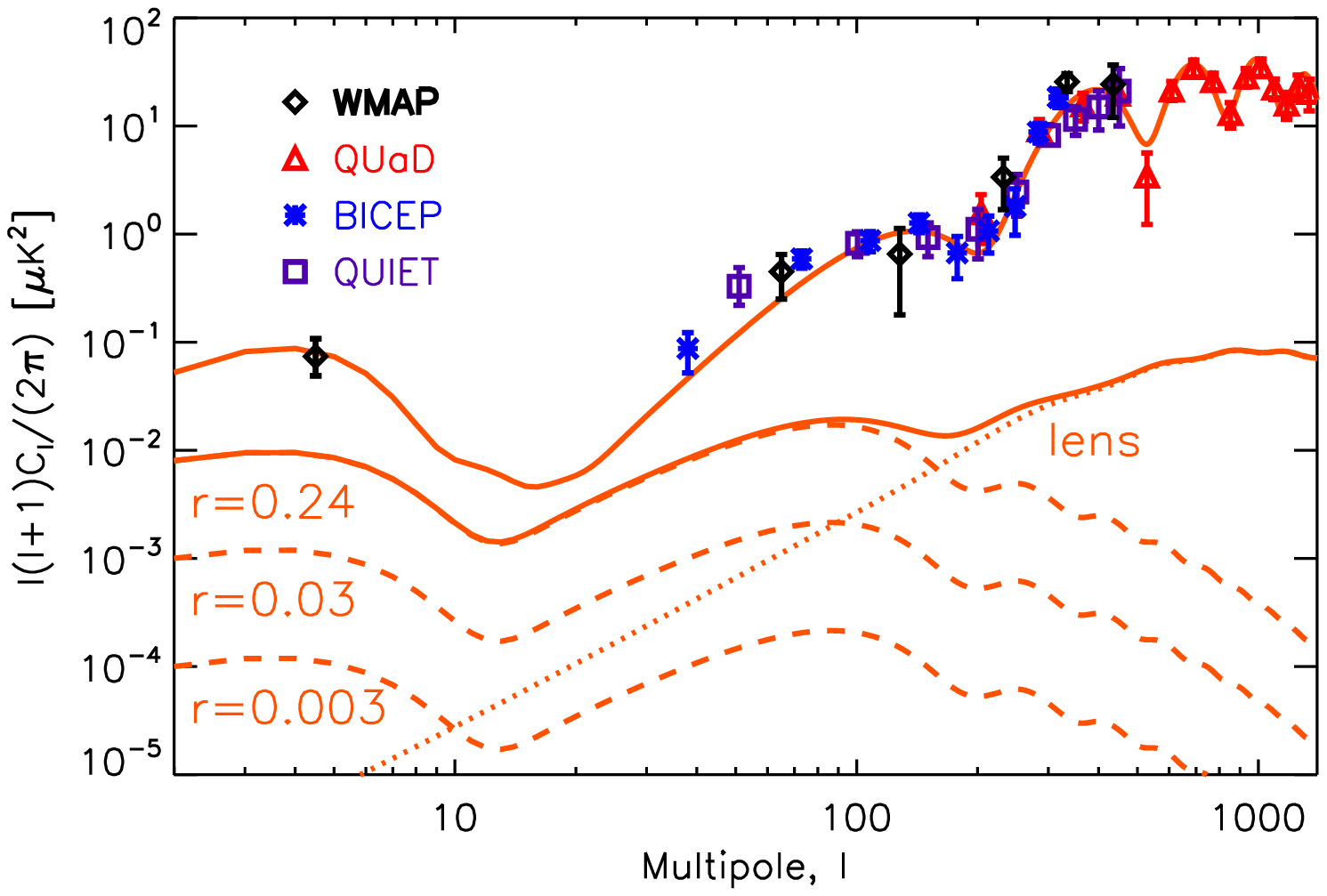}
\caption{%
$E$-mode and $B$-mode polarization power spectra. The diamonds,
 triangles, stars, and squares show the {\sl WMAP} seven-year data
 \citep{larson/etal:prep}, the QUaD final data \citep{brown/etal:2009},
 the BICEP two-year data \citep{chiang/etal:2010}, and the QUIET 43~GHz
 data \citep{quiet:prep}, respectively. The upper solid line shows the
 scalar $E$-mode power spectrum of the {\sl WMAP} seven-year best-fit
 model. The dashed lines show the primordial $B$-mode power spectra with
 the tensor-to-scalar ratio of $r=0.24$, which corresponds to the
 current 95\% upper limit \citep{komatsu/etal:prep}, as well as of
 $r=0.03$ and 0.003. These lines are linearly proportional to $r$. The
 dotted line shows the secondary $B$-mode power spectrum expected to be
 generated by the weak gravitational lensing effect converting $E$ modes
 to $B$ modes \citep{zaldarriaga/seljak:1998}. This line is fixed (by
 the {\sl WMAP} seven-year best-fit model) and acts as noise for the
 primordial $B$-mode detection. The lensing contribution becomes
 comparable to the primordial bump at $l=10$ and 100 for $r=0.003$ and
 0.03, respectively.
} 
\label{fig:clnow}
\end{figure}

Given the upper limit on $r$, one can calculate the expected level of
the $B$-mode power spectrum (see Figure~\ref{fig:clnow}). For $r=0.24$, the
$B$-mode power spectrum is smaller than the $E$-mode power spectrum by a
factor of 10 at the first bump (created by electrons at $z\lesssim
10$). At the second bump (created by electrons at $z\simeq 1090$), the
$B$-mode power spectrum is smaller than the $E$-mode power spectrum by a
factor of 50. It is the smallness  of the $B$-mode power spectrum that
makes the detection of this signal challenging. 

There are three sources of noise for $B$-mode detection:
(1) Detector noise; (2) Galactic foreground emission; and (3)
Gravitational lensing. In this paper, we shall focus on the Galactic
foreground. We use a map-based method for
reducing the Galactic foreground, and study how the residual foreground
limits a measurement of the primordial $B$-mode polarization. The
foreground reduction technique we use is motivated by 
the ``template cleaning method'' used by the {\sl WMAP} team
\citep{page/etal:2007,gold/etal:2009,gold/etal:prep}. This method was
further investigated by \citet{efstathiou/gratton/paci:2009} in the
context of the {\sl Planck} mission.
We shall study this technique in the context of 
a next-generation, low-noise, low-resolution (0.5 degree FWHM) space-borne
experiment. 

There is a large body of literature on the issue of polarized foreground
cleaning for the $B$-mode detection. Our method is one specific
(and relatively simpler) example. For the other methods in the literature, see
review articles \citep{dunkley/etal:2008,fraisse/etal:prep} and
references therein. 

This paper is organized as follows.
In Section~\ref{sec:noise}, we show how the detector noise and the
lensing noise influence the statistical errors on $r$. 
In Section~\ref{sec:method}, we describe our method for estimating
$r$ in the presence of the Galactic foreground and the dominant scalar
$E$-mode polarization. In Section~\ref{sec:simulation}, we describe our
simulation including CMB, detector noise, and foreground.
In Section~\ref{sec:results}, we present the main results of this
paper. We conclude in Section~\ref{sec:conclusion}.

\section{Detector noise and lensing noise}
\label{sec:noise}
Before we study the effect of the foreground, we show how the detector
noise and the lensing noise influence our ability to detect $r$. The
detector noise enters 
into the likelihood of $r$ via the noise power spectrum, $N_l^{BB}$. Assuming
white noise, we write the noise power spectrum as
\begin{equation}
 N_l^{BB} = \left(\frac{\pi}{10800}\frac{{w}_p^{-1/2}}{{\mu{\rm K}}~\rm
	    arcmin}\right)^2\mu{\rm K}^2~{\rm str},
\label{eq:nlbb}
\end{equation}
where ${w}_p^{-1/2}$ is the noise in Stokes parameters $Q$ or $U$ per
pixel whose solid angle, $\Omega_{\rm pix}$, gives $\sqrt{\Omega_{\rm
pix}}=1$~arcmin. This quantity is useful because one can compare various
experiments on the same scale.

Current and future experiments use many (of order $10^3-10^4$)
detectors to reduce the noise equivalent temperature (NET) down to a few
$\mu$K~arcmin level. Is this sufficient for detecting primordial $B$
modes? For comparison, the expected sensitivity of {\sl Planck} combining
70, 100, and 143~GHz is ${w}_p^{-1/2}=63~{\mu}$K~arcmin \citep[see,
e.g., Appendix A of][]{zaldarriaga/etal:prep, planck:bb}. 

\begin{figure}[t]
\centering \noindent
\includegraphics[width=8.5cm]{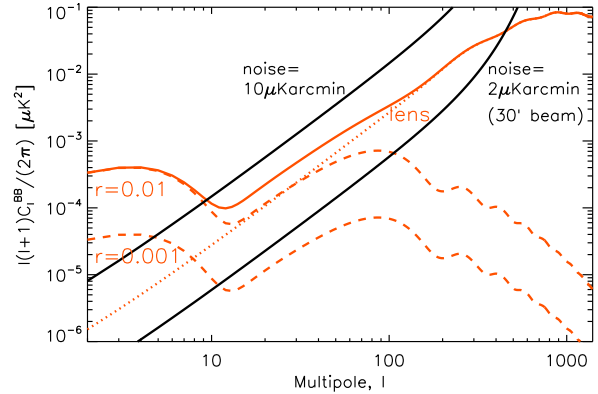}
\caption{%
 $B$-mode polarization signal and noise power spectra. The dashed lines
 show the primordial $B$-mode power spectra with the tensor-to-scalar
 ratio of $r=0.01$ and 0.001, while the dotted line shows the secondary
 $B$-mode power spectrum from gravitational lensing. We also show the
 noise power spectra (Equation~(\ref{eq:nlbb})) for $w_p^{-1/2}=2$ and
 10~$\mu$K~arcmin with a Gaussian beam window function of $\theta_{\rm
 FWHM}=30$~arcmin, i.e., $N_l^{BB}e^{l^2\theta_{\rm FWHM}^2/(8\ln 2)}$.
} 
\label{fig:clnoise}
\end{figure}

In Figure~\ref{fig:clnoise}, we compare the noise power spectra
for $w_p^{-1/2}=2$ and 10~$\mu$K~arcmin to the primordial and lensing
$B$ modes. For $r=10^{-3}$ and the 10~$\mu$K~arcmin noise, only a few
modes ($l=2$, 3, and 4) are above noise. For the 2~$\mu$K~arcmin noise,
the noise power spectrum is below the lensing $B$-mode power spectrum,
and thus noise is no longer the limiting factor (unless we ``de-lens''
maps and remove the lensing noise). How would this influence our ability
to detect $r$? 

\begin{figure}[t]
\begin{center}
\includegraphics[width=8.5cm]{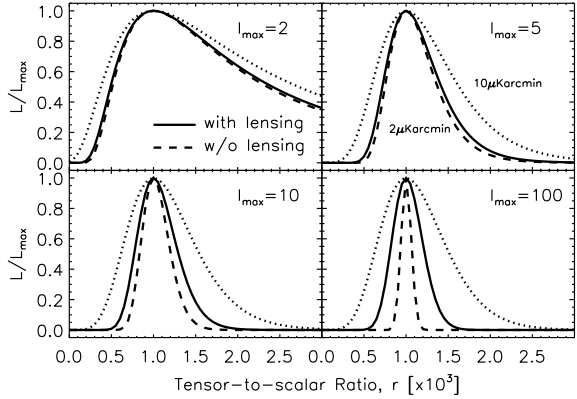}
\caption{%
 Effect of detector noise and gravitational lensing on the likelihood
 for $r$ (foreground is not included). The input value of $r$ is $r_{\rm
 input}=10^{-3}$. The values on the horizontal axis have been multiplied
 by $10^3$. In the top left, top right, bottom left, and bottom right
 panels, we sum the multipoles up to $l_{\rm max}=2$, 5, 10, and 100,
 respectively. For the detector noise level, we use $2~{\mu}$K~arcmin
 for the solid and dashed lines, and $10~{\mu}$K~arcmin for the dotted
 lines. The solid and dotted lines include the gravitational lensing
 contribution to the total noise, while the dashed lines do not. Even if
 we set the detector noise to be zero, the solid lines do not change
 very much: for $r=10^{-3}$, the gravitational lensing effect prevents
 us from measuring $r$ accurately beyond $l\sim 10$. Note that a single
 multipole, $l=2$, is enough for us to detect $r=10^{-3}$ if the
 detector noise is smaller than $10~{\mu}$K~arcmin.
}
\label{fig:like}
\end{center}
\end{figure}

\begin{figure}[t]
\begin{center}
\includegraphics[width=8.5cm]{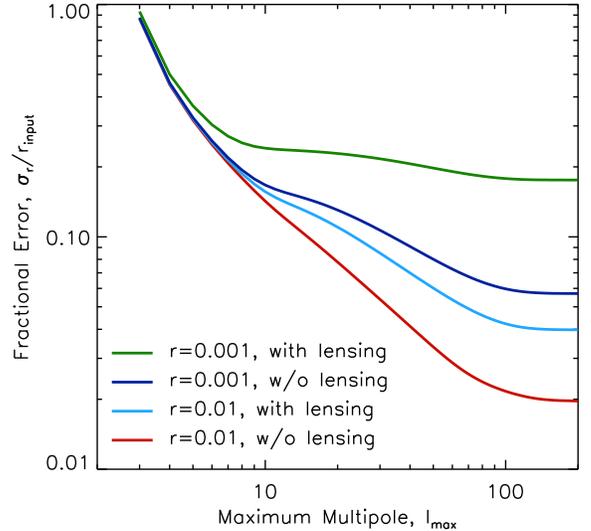}
\caption{%
 Fractional error, $\sigma_r/r_{\rm input}$, on the determination of the
 value of  $r$  as a function of maximum multipoles, $l_{\rm
 max}$. Here, $\sigma_r$ is the square-root of the second-order moment
 of the likelihood 
 function given by  Equation~(\ref{eq:sigma}).
 (The foreground is not included. The full sky coverage is assumed.)
 From the top to bottom lines, we show $r_{\rm input}=0.001$ with and
 without the lensing noise, and $r_{\rm input}=0.01$ with and
 without the lensing noise, respectively. For the instrumental noise level, we
 have used 2~${\mu}$K~arcmin. Note that $\sigma_r/r_{\rm input}\sim 1$
 at $l_{\rm max}=3$ does not mean that we do not detect $r$; on the contrary, we
 detect $r$ with high significance even at $l_{\rm max}=2$. Rather, it
 just means that the likelihood for $r$ is highly non-Gaussian and has a
 long tail toward large values of $r$ (see the top left panel of
 Figure~\ref{fig:like}). 
 In other words, we detect $r$ with high statistical significance, 
 but the value of $r$ is not determined very well.
}
\label{fig:significance}
\end{center}
\end{figure}

To see this, let us calculate the likelihood of $r$ for a given noise
level. For simplicity, we assume that we cover the full sky and the
noise per pixel is homogeneous.\footnote{We assume this only for
producing Figures~\ref{fig:like} and \ref{fig:significance}. For the
main analysis, we include inhomogeneous noise, foreground, and a partial
sky coverage.} Then, one can write down the
probability distribution function of the measured $B$-mode power
spectrum, $\hat{C}_l^{BB}$, for a given value of $r$ as \citep[e.g.,
Equation~(8) of][]{hamimeche/lewis:2008} 
\begin{eqnarray}
\nonumber
& & -2\ln
  P(\hat{C}^{BB}_l|r)\\
\nonumber
& =&(2l+1)\left[\frac{\hat{C}^{BB}_l}{rc_l^{GW}+c_l^L+N_l^{BB}}
+\ln(rc_l^{GW}+c_l^L+N_l^{BB})
\right.\\
 & &\left.\qquad\qquad
-\frac{2l-1}{2l+1}\ln(\hat{C}_l^{BB})
\right],
\end{eqnarray}
where $c_l^{GW}$ is the primordial $B$-mode power spectrum from
gravitational waves with $r=1$, and $c_l^L$ is the secondary $B$ mode
from gravitational lensing.
We then use Bayes' theorem to calculate the likelihood for $r$ as
${\cal L}(r|\hat{C}^{BB}_l)\propto P(\hat{C}^{BB}_l|r)$.
To calculate the likelihood, we set the measured power spectrum
to be $\hat{C}^{BB}_l=r_{\rm input}c_l^{GW}+c_l^L+N_l^{BB}$, and sum the
log-likelihood over multipoles up to $l_{\rm max}$: 
\begin{equation}
 \ln{\cal L}(r)=\sum_{l=2}^{l_{\rm max}}\ln{\cal L}(r|\hat{C}^{BB}_l).
\end{equation}

Figure~\ref{fig:like} shows the likelihood of $r$ for the input
value of $r_{\rm input}=10^{-3}$ and $l_{\rm
max}=2$, 5, 10, and 100. One useful number to keep in mind is that a
{\it single} multipole, $l=2$, is sufficient for detecting $r=10^{-3}$,
if the noise is smaller than $10~{\mu}$K~arcmin. However, the precision
on $r$ does not improve beyond $l=5$. This is apparent also in
Figure~\ref{fig:clnoise}: the noise power spectrum exceeds the signal
at $l\ge 5$.

We can improve the precision further if we lower the noise level to,
say, $2~{\mu}$K~arcmin. Even so, the gravitational lensing prevents us
from improving on the precision beyond $l\sim 10$ if $r=10^{-3}$. (If
there were no lensing in the universe, we would be able to continue to
improve on the precision, as indicated by the dashed lines.) In fact,
$2~{\mu}$K~arcmin is essentially the same as zero detector noise, as the
lensing term dominates the error budget. Again, this is apparent in
Figure~\ref{fig:clnoise}.

Of course, these results are overly
optimistic, as the error would be dominated by the foreground rather
than by the detector noise. Nevertheless, it is still useful to know 
what would be possible when we ignore the foreground.

To quantify the precision on $r$, it is convenient to use the variance,
$\sigma^2_r$, given by the second moment of the likelihood:
\begin{equation}
\label{eq:sigma}
 \sigma^2_r = \int_0^\infty dr {\cal L}(r)r^2 
- \left[\int_0^\infty dr {\cal L}(r)r\right]^2.
\end{equation}
Here, we have assumed that the likelihood is normalized such that
$\int_0^\infty dr {\cal L}(r)=1$.
One should be careful when interpreting this quantity. For $l_{\rm
max}=2$, $\sigma_r$ would be greater than the input value, $r_{\rm
input}=10^{-3}$; however, this does not mean that we cannot
detect $r$. This just means that the distribution is highly non-Gaussian
and has a long tail toward large values of $r$ (see the top left panel
of Figure~\ref{fig:like}). For large values of $l_{\rm max}$,
e.g., $l_{\rm max}\gtrsim 10$, the distribution of $r$ becomes approximately a
Gaussian, and thus the value of $\sigma_r$ may be interpreted as the size of
the usual $1\sigma$ error bar.

Figure~\ref{fig:significance} shows 
the fractional error, $\sigma_r/r_{\rm
input}$, on the determination of the value of $r$
as a function of $l_{\rm
max}$. First, as one may 
expect from Figure~\ref{fig:like}, the fractional error for $r_{\rm
input}=10^{-3}$ saturates 
at $l_{\rm max}\sim 10$ and does not improve further due to the lensing
noise. For this case, while we can detect $r$ with high statistical
significance,
we can determine the actual value of $r$ to only $\sim 20$\%. 
For $r_{\rm
input}=10^{-2}$, we can determine the value of $r$ to $\sim 4$\% at
$l_{\rm max}=200$ (beyond which the fractional error no longer improves
due to the 
lensing noise).

This study gives us an estimate of statistical errors on the
measured 
values of $r$. On the other hand, the Galactic foreground gives us 
systematic errors (and bias). Now we shall turn to the foreground
issue, which is the main subject of this paper.
\section{Pixel-based foreground removal method}
\label{sec:method}
\subsection{Motivation}
The basic idea behind our methodology is simple: 
we have (at least) 3 polarized components on the sky that we know and 
have been detected by the {\sl WMAP}: CMB, synchrotron emission, and thermal
dust emission. As the synchrotron dominates at lower frequencies and the
dust at higher frequencies, we use one map at a low frequency and
another map at a high frequency as the foreground ``templates.''
We put the quotation marks here because these maps also contain the
CMB. No {\it external} template maps are used in our method.

The {\sl WMAP} team has applied this method for modeling the
synchrotron: they used the lowest frequency (K-band, 23~GHz) map as a
template, fitted it to the higher frequency maps (Ka, Q, V, and W bands),
and subtracted from those maps. One can write this operation as
\begin{eqnarray}
\label{eq:quclean}
 [Q',U'](\nu)=\frac{[Q,U](\nu)-\alpha_S(\nu)[Q,U](\nu=23~{\rm GHz})}{1-\alpha_S(\nu)},
\end{eqnarray}
where $Q'$ and $U'$ are the template-cleaned Stokes $Q$ and $U$ maps,
respectively, and $\alpha_S$ is the best-fit synchrotron coefficient for
a given frequency $\nu$. The denominator accounts for the fact that the
K-band map also contains the CMB signal.

However, the {\sl WMAP} team had to rely on an external map for modeling
the dust emission, as the highest frequency, the W band (94~GHz), was
not high enough for being a good template of the polarized dust
emission. This issue would probably be resolved by the {\sl Planck}
satellite, which has higher frequency channels such as 217 and 353~GHz. 
 \citet{efstathiou/gratton/paci:2009} have studied this by 
using a simulated {\sl Planck} 217~GHz or 353~GHz map as a template for
dust, and a simulated 30~GHz map as a template for synchrotron.
They find that this simple method removes the foreground efficiently,
bringing the 
bias in $r$ down to a few times $10^{-3}$, which is much smaller than
the expected statistical uncertainty on $r$ from {\sl Planck},
$\sigma_r={\cal O}(10^{-2})$.

The goal of this paper is to put this method in the context of a
next-generation, low noise (2~$\mu$K~arcmin) polarization satellite
experiment, and see if this method yields a promising result for
measuring $r\sim 10^{-3}$ (which is easy to detect in the absence of 
foreground, as we just saw in Section~\ref{sec:noise}).

\subsection{``Template'' cleaning method}
Our methodology is similar to that given in Section~4.2 of
\citet{efstathiou/gratton/paci:2009}. 

The main parameter that we wish to extract from data is 
the tensor-to-scalar ratio, $r$. (We do not vary the tensor
tilt, $n_t$.) The foreground coefficients,
$\alpha$, are nuisance parameters that we wish to marginalize over.
The foreground coefficients may be spatially varying.

Another nuisance parameter (for detecting $B$ modes) is the amplitude of
the scalar $E$-mode power spectrum, which is by far the dominant
source of CMB polarization. 
The signal power spectra are thus given as
\begin{eqnarray}
 C_l^{EE} &=& sc_l^{{\rm scalar},EE}+rc_l^{{\rm tensor},EE},\\
 C_l^{BB} &=& rc_l^{{\rm tensor},BB},
\end{eqnarray}
where $c_l$ denotes the power spectra with $s=1$ and $r=1$. The fiducial
value of $s$ is $s=1$.

We shall maximize the following likelihood function for estimating $r$, $s$,
and $\alpha_i$:
\begin{equation}
 \mathcal{L}(r, s, \alpha_i) \propto \frac{\exp\left[-\frac{1}{2}{\bm x'(\alpha_i)}^T{\bm C^{-1}(r, s,
  \alpha_i)}{\bm x'(\alpha_i)}\right]}{\sqrt{\vert {\bm
  C(r, s, \alpha_i)}\vert}},
\label{eq:fulllike}
\end{equation}
where 
\begin{equation}
\label{eq:cleanedmap}
\bm{x}'=\frac{[Q,U](\nu)-\sum_i\alpha_i(\nu)[Q,U](\nu^{\rm
 template}_i)}{1-\sum_i\alpha_i(\nu)} 
\end{equation}
 is a template-cleaned map. This is a generalization of
 Equation~(\ref{eq:quclean}) for a multi-component case. In this paper,
 $i$ takes on ``S'' and ``D'' for 
synchrotron and dust, respectively, unless noted otherwise. For
 definiteness, we shall choose: 
\begin{eqnarray*}
 \nu = \mbox{100 GHz},\\
 \nu^{\rm template}_{\rm S} = \mbox{60 GHz},\\
 \nu^{\rm template}_{\rm D} = \mbox{240 GHz}.
\end{eqnarray*}
These choices are somewhat arbitrary, but our preliminary optimization
study indicates that this is a good configuration for achieving
a smaller bias in $r$. A fuller optimization study, including more
frequency channels, would require a more
detailed specification of a given experiment (e.g., how many detectors
one can fit in a given focal place; how low the detector noise can be as
a function of frequencies), which is beyond the scope
of this paper, but will be presented elsewhere. 

The covariance matrix in pixel space, ${\bm C}$, for Stokes $Q$ and $U$ maps is
given as 
\begin{equation}
\label{eq:cleanedcinv}
{\bm C}(r,s,\alpha_i) = r {\bm c}^{\rm tensor} + s {\bm c}^{\rm
 scalar} + \frac{{\bm N}_1 + {\bm N}_2}{(1-\sum_i\alpha_i)^2},
\end{equation}
where ${\bm c}$ is the signal covariance matrix calculated from the
theoretical power spectra, $c_l$, (see Appendix~\ref{sec:cov}) and the
noise matrices, 
${\bm N}_1$ and ${\bm N}_2$, are a noise covariance of a smoothed map
(which is not diagonal)
{\it before the template cleaning is applied},
 and a small artificial diagonal noise
 matrix for a matrix regularization, respectively
(see Section~\ref{sec:cmbnoise} for details).

For simplicity and clarity, we have ignored noise in template maps.
For, if we assume that all three
channels are similar in detector noise level, it
is a good
approximation, as $\alpha_{\rm D}\sim 0.08$ and
$\alpha_{\rm S}\sim 0.25$, and the fractional 
contribution of the template noise to the covariance matrix is given by
$\alpha^2_i$, i.e., 6\% effect in the derived error bars. Note
that this is equivalent to ignoring {\bf P} in Section~4.2 of
\citet{efstathiou/gratton/paci:2009}.
\section{Simulation}
\label{sec:simulation}
\subsection{CMB and detector noise}
\label{sec:cmbnoise}

\begin{figure*}[t]
\begin{center}
\includegraphics[width=15cm]{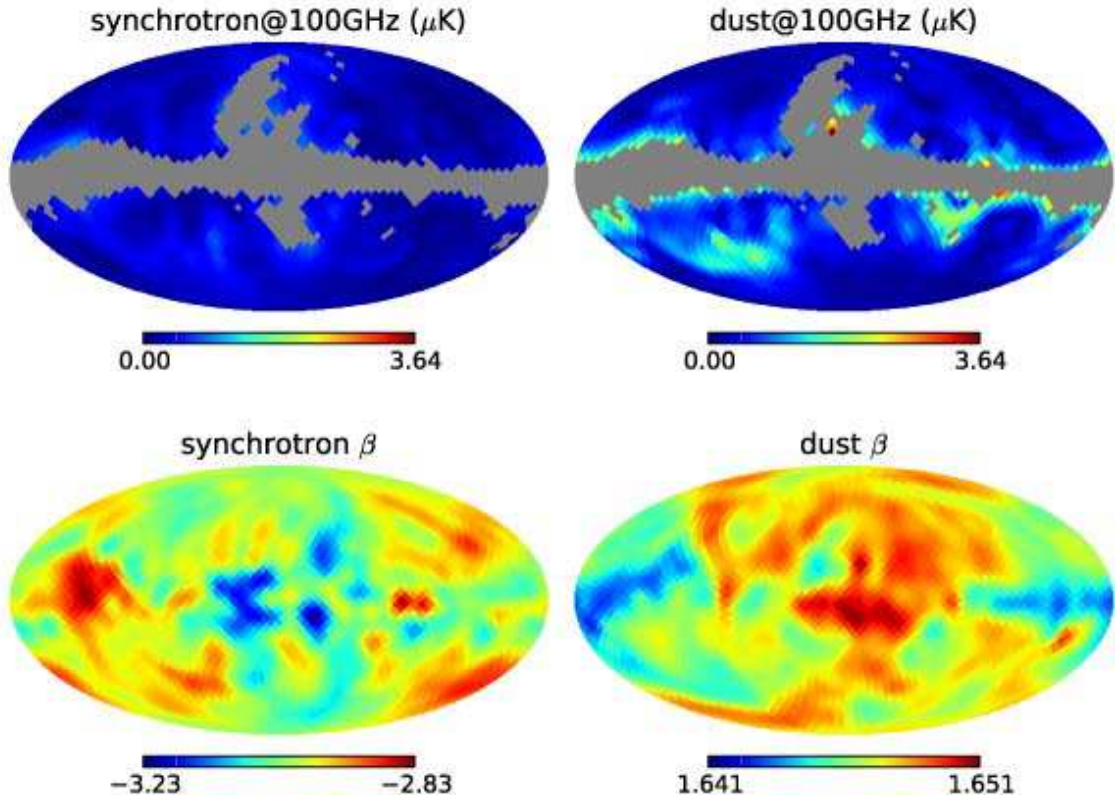}
\caption{%
 Foreground maps from the Planck Sky Model (PSM; v1.6.2). The top left
 and top right panels show the polarization intensity maps
 ($P=\sqrt{Q^2+U^2}$ in units of $\mu$K) of synchrotron and dust,
 respectively. The dust polarization intensity has been multiplied by
 a factor of three to better approximate a more recent version of PSM.
 The lower left and lower right panels show the
 synchrotron index $\beta_S$ and the dust index $\beta_D$,
 respectively. Note a small range shown for $\beta_D$: the dust index
 does not vary much, but this is a built-in assumption of the PSM
 v1.6.2.
}
\label{fig:foregrounds}
\end{center}
\end{figure*}

For CMB, we first generate the scalar and tensor polarization
power spectra using the CAMB code \citep{lewis/challinor/lasenby:2000}
with and without lensing contributions. 
We then generate Stokes $Q$ and $U$ maps at the Healpix resolution of
$N_{\rm side}=128$. The signal map has been smoothed with a $30'$ beam
(FWHM), representing a low-angular-resolution CMB polarization satellite
experiment targeting the primordial $B$ modes.

To this smoothed signal map, we add random Gaussian noise given by 
$\sigma_0/\sqrt{N_{\rm obs}(\hat{\bm n})}$ per pixel in the direction of
$\hat{\bm n}$. Here, $\sigma_0$ is related to noise $w_p^{-1/2}$ as
\begin{equation}
 \sigma_0 = \frac{\pi}{10800}\frac{w_p^{-1/2}}{{\mu}{\rm
  K~arcmin}}\frac1{\sqrt{\Omega_{\rm pix}N_{\rm
  pix}^{-1}\sum_iN^{-1}_{\rm obs}(\hat{\bm n}_i)}}~\mu{\rm K},
\end{equation}
where $N_{\rm pix}=12(128)^2=196608$ is the total number of pixels at
$N_{\rm side}=128$, and $N_{\rm obs}$ is the number of observations per
pixel. We adopt $N_{\rm obs}$ from the ``EPIC low-cost'' (EPIC-LC)
design \citep{bock/etal:prep}. The noise is
highest on the ecliptic plane and lowest on the ecliptic poles, similar
to the $N_{\rm obs}$  pattern of the {\sl WMAP}. Note that the absolute
value of $N_{\rm obs}$ will cancel out in $\sigma_0/\sqrt{N_{\rm
obs}(\hat{\bm n})}$ if we use the above formula: only the spatial
distribution is taken from $N_{\rm obs}$, and the overall noise level
is set by the assumed value of $w_p^{-1/2}$. 
We shall use $w_p^{-1/2}=2~\mu$K~arcmin for the rest of this paper.
For this low noise configuration, the results are not sensitive to the
details of the $N_{\rm obs}$ pattern.

As we described at the end of Section~\ref{sec:method}, noise in
template maps (at $\nu_{\rm S}=60$~GHz and $\nu_{\rm D}=240$~GHz) makes
only a small contribution to the final covariance matrix. Therefore, for
simplicity we add noise only to our CMB channel at
100~GHz.\footnote{Note that noise in templates cannot be
ignored when we try to find an optimal combination of 3 frequencies. We ignore
noise in templates here because we have done our preliminary
optimization already. A fuller exploration of template noise along with
the frequency optimization will be given elsewhere.}

We then apply an additional Gaussian smoothing to this signal-plus-noise
map with 9.16 degrees (FWHM), which is 
$2.5$ times the pixel size at $N_{\rm side}=16$, and re-sample the
smoothed map to $N_{\rm side}=16$.
Finally, as the smoothed map at $N_{\rm side}=16$ is dominated by the
scalar $E$-mode signal
at all angular scales supported by the map resolution, the covariance
matrix of this map is singular. In order to regularize the covariance matrix, 
we add an artificial, homogeneous white noise of $0.2~\mu{\rm
K~arcmin}$ such that the map becomes
noise dominated at the Nyquist frequency, $l_{\rm max}=3N_{\rm nside}-1=47$.

\subsection{Foreground: Planck Sky Model}
\begin{figure}[t]
\centering \noindent
\includegraphics[width=8.5cm]{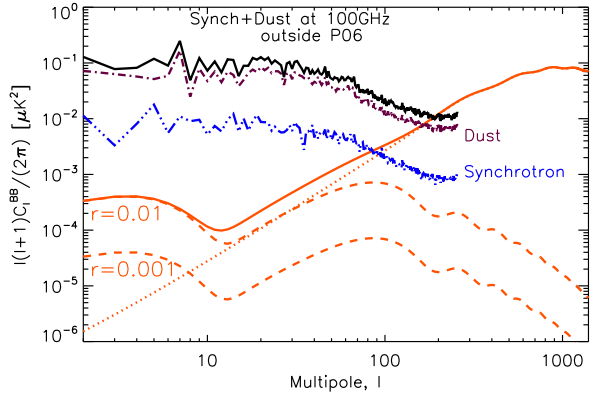}
\caption{%
 $B$-mode polarization signal and foreground power spectra. The dashed
 lines show the primordial $B$-mode power spectra with the
 tensor-to-scalar ratio of $r=0.01$ and 0.001, while the dotted line
 shows the secondary $B$-mode power spectrum from gravitational
 lensing. We also show the $B$-mode power spectra of the dust
 (dot-dashed line) and synchrotron (triple-dot-dashed line) emission at
 100~GHz outside the {\sl WMAP} P06 mask ($N_{\rm side}=128$). (The
 measured foreground power 
 spectra have been divided by $f_{\rm sky}=0.733$ to approximately
 correct for the mask, as well as by the pixel window
 function at $N_{\rm side}=128$.) The total (synch$+$dust)
 power spectrum is 
 measured from the total {\it map}, which is slightly larger than the
 sum of the synchrotron and dust power spectra, as these two foreground
 components are spatially correlated. Note that the original PSM v1.6.2 dust map
 has the average polarization fraction of 1.5\%, but we
 have multiplied the dust map by a factor of 3 to approximate a more recent
 dust template map adopted by the Planck collaboration.
} 
\label{fig:clfg}
\end{figure}

For the Galactic foreground model, we use the Planck Sky Model (PSM; v1.6.2) 
developed by the Planck Component Separation Working Group (Working
Group 2). \citet{leach/etal:2008} describe the PSM for temperature, and
\citet{dunkley/etal:2008} for polarization.

The polarized synchrotron and dust emission are modeled as power-laws in
antenna temperature: 
\begin{eqnarray}
\nonumber
[Q_{\rm synch},U_{\rm synch}](\nu,\hat{\bm n})
 &=& g(\nu)[\tilde{Q}^{\rm PSM}_{\rm synch},\tilde{U}^{\rm PSM}_{\rm synch}](30~{\rm
 GHz},\hat{\bm n}), \\
\label{eq:synch}
& &\times \left(\frac{\nu}{30~{\rm GHz}}\right)^{\beta_{\rm S}(\hat{\bm n})}, \\ \nonumber
[Q_{\rm dust},U_{\rm dust}](\nu,\hat{\bm n})
 &=& g(\nu)[\tilde{Q}^{\rm PSM}_{\rm dust},\tilde{U}^{\rm PSM}_{\rm dust}](94~{\rm
 GHz},\hat{\bm n}), \\
& &\times \left(\frac{\nu}{94~{\rm GHz}}\right)^{\beta_{\rm D}(\hat{\bm n})}.
\label{eq:dust}
\end{eqnarray}
Here, $\tilde{Q}^{\rm PSM}$ and $\tilde{U}^{\rm PSM}$ are the PSM Stokes
parameters in units of  
antenna temperature, and $g(\nu)\equiv (e^x-1)^2/(x^2e^x)$ where
$x=h\nu/k_BT_{\rm CMB}=\nu/56.780$~GHz converts the antenna temperature
to thermodynamic temperature. (${Q}$ and ${U}$ are in units of
thermodynamic temperature.) 

For synchrotron, the position-dependent spectral index,
$\beta_{\rm S}(\hat{\bm n})$, is calculated from the Haslam 408~MHz map
\citep{haslam/etal:1981} and the three-year {\sl WMAP} temperature map
at 23~GHz \citep{page/etal:2007}. The template maps at 30~GHz are taken from
\citet{miville-deschenes/etal:2008}.

For dust, the position-dependent spectral index, $\beta_{\rm D}(\hat{\bm
n})$, as well as the unpolarized intensity map are taken from 
Model 8 of \citet{finkbeiner/davis/schlegel:1999}. 
The polarization angles of dust approximately follow those of the
synchrotron maps. The original PSM dust map has the average polarization
fraction of 1.5\% over the full sky, but we will multiply this map by
a factor of 3 to approximate a more recent dust map used by the Planck
collaboration.

Top panels of Figure~\ref{fig:foregrounds} show the amplitude
($P=\sqrt{Q^2+U^2}$) of polarization intensity of synchrotron and dust at
100~GHz, while the bottom panels show the spectral indices, $\beta_{\rm
S}$ and $\beta_{\rm D}$. After adding the above foreground maps
(smoothed with a 9.16-degree beam at $N_{\rm side}=128$ and degraded to
$N_{\rm side=16}$) to the CMB-plus-noise map, we mask the simulated sky
by the {\sl WMAP} P06 mask ($f_{sky}=73\%$)\citep{page/etal:2007}.

The norm of the pixel vector, [$Q$,$U$], is $2259\times 2$, where 2259
is the number of pixels outside the P06 mask. In order to mask the
covariance matrix, we use the technique
described in Appendix D of \citet{page/etal:2007}: we compute an inverse
of $6144\times 6144$ matrix and reduce it to $4518\times 4518$ matrix using
Equation~(D7) of \citet{page/etal:2007}. (Note that there is a typo in
this equation: $D$ should be replaced by $D^{-1}$.)

In Figure~\ref{fig:clfg}, we show the $B$-mode power spectra measured
from the PSM ($N_{\rm side}=128$) at 100~GHz outside the P06 mask. 
The total foreground power spectrum has $l(l+1)C_l^{BB}/(2\pi)\approx
10^{-1}~\mu{\rm K}^2$ at $l\lesssim 10$, which is 250 and 2500
times larger than 
the primordial $B$-mode spectra with $r=0.01$ and 0.001, respectively.
The problem seems formidable; however, as we show below, the simple cleaning
method can reduce the foreground-induced bias in $r$ to $\Delta r\approx
0.002(<0.001)$ with the P06(extended) mask.
\section{Results}
\label{sec:results}
\subsection{Fixing the scalar $E$-mode amplitude}
\label{sec:offset}
\begin{figure*}[t]
\begin{center}
\includegraphics[width=8cm]{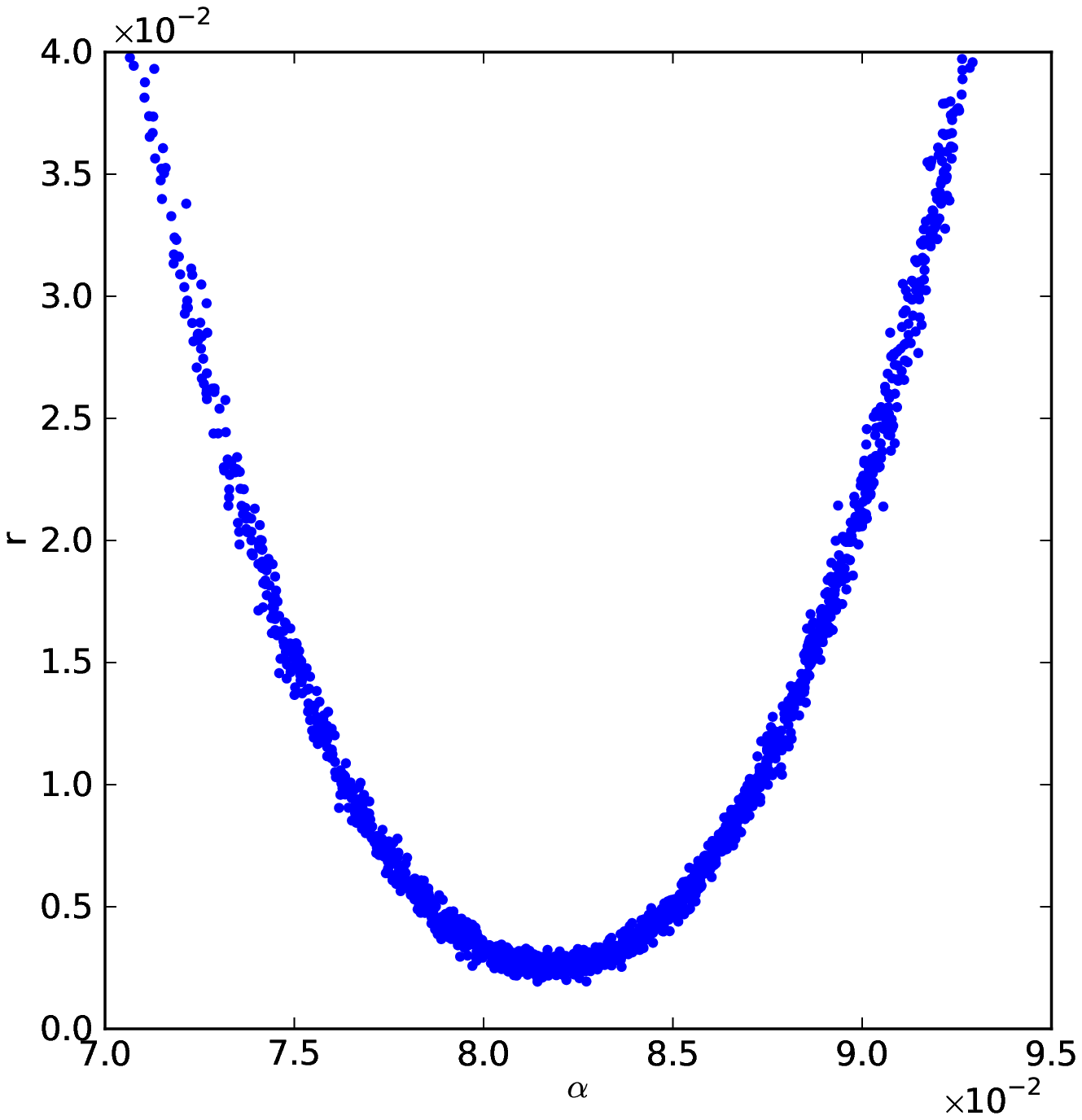}
\includegraphics[width=8cm]{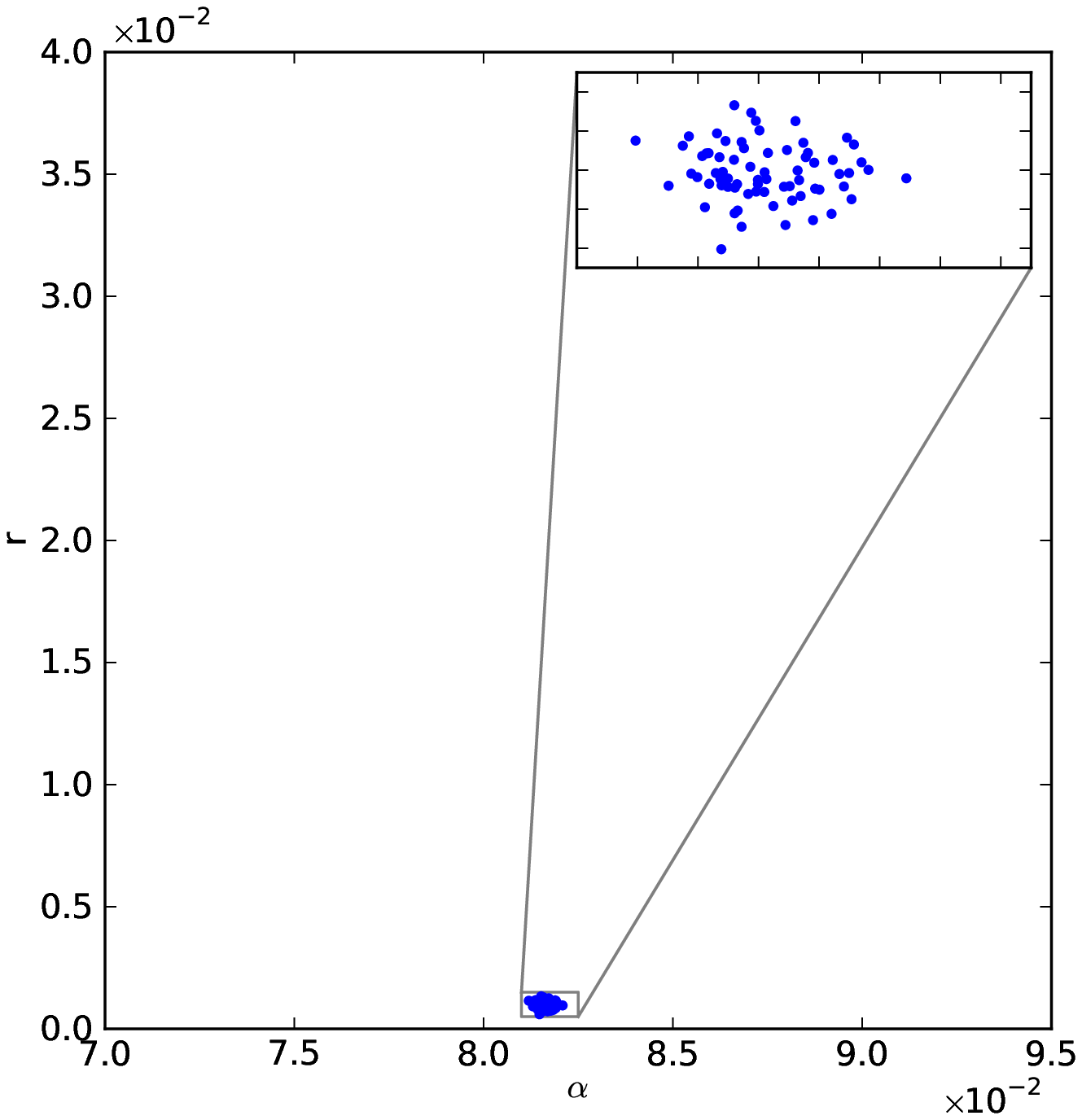}
\caption{%
 Correlation between the tensor-to-scalar ratio, $r$, and the amplitude
 of dust, $\alpha_{\rm D}$. Left: The amplitude of the scalar $E$ modes
 is held fixed at $s=1$. Right: The amplitude of the scalar $E$ modes is
 treated as a nuisance parameter and marginalized over. The input
 tensor-to-scalar ratio is $r_{\rm input}=0.003$ and, for this figure only,
 the original PSM dust map (with an average polarization of $\sim
 1.5\%$) is used.
}
\label{fig:r-alpha}
\end{center}
\end{figure*}

Before we use our full likelihood function given by
Equation~(\ref{eq:fulllike}), let us first try a simpler version and
show that it actually fails. 

For the moment (only within this subsection), we fix the amplitude of 
the scalar $E$ modes, i.e., $s=1$, and consider cleaning dust using a
map at 240~GHz. (Synchrotron will not be discussed in this subsection.)
Our model is thus
\begin{eqnarray}
\left[Q,U\right](100) &=& \mbox{CMB} + \mbox{Dust(100)} + \mbox{Noise}, \\
\left[Q,U\right](240) &=& \mbox{CMB} + \mbox{Dust(240)}.
\end{eqnarray}
As we described at the end of Section~\ref{sec:method}, we ignore noise
at 240~GHz. We then fit the 240~GHz map to the 100~GHz map:
\begin{equation}
[Q',U'](100) = [Q,U](100) - \alpha_{\rm D}[Q,U](240).
\end{equation}
Minimizing 
$\chi^2 = [Q',U']^T{\bm C}^{-1}[Q',U']$ with respect to $\alpha_{\rm D}$
gives the following least-square solution:
\begin{equation}
\alpha_{\rm D} = \frac{[Q,U]^T(100){\bm C}^{-1}[Q,U](240)}{[Q,U]^T(240){\bm C}^{-1}[Q,U](240)}.
\end{equation}
As the polarization signal is dominated by scalar $E$ modes, we can set $r=0$
when computing the covariance matrix ${\bm C}$ in this equation.
(In practice, we used $r_{\rm input}$.)
Finally, we maximize the likelihood given in
Equation~(\ref{eq:fulllike}) with respect to $r$, with $s=1$ and
$\alpha_{\rm D}$ given by the above least-square solution.

The left panel of Figure~\ref{fig:r-alpha} shows the values of $r$ and
$\alpha_{\rm D}$ obtained from many random realizations of noise and CMB skies.
(The input tensor-to-scalar ratio is $r_{\rm input}=0.003$.)
There is a clear correlation between $r$ and $\alpha_{\rm D}$,
indicating a failure of this algorithm. This correlation is caused by a
chance correlation between foreground and the dominant scalar $E$ modes
\citep{chiang/naselsky/coles:2008,efstathiou/gratton/paci:2009}. The
correlation disappears when we set $C_l^{EE}=0$. This result motivates our
treating 
the amplitude of scalar modes as a nuisance parameter.

The right panel of Figure~\ref{fig:r-alpha} shows the results when $s$
is treated as a nuisance parameter and marginalized over. For this, we
have maximized the likelihood given by Equation~(\ref{eq:fulllike}) by
varying $r$, $s$, and $\alpha_{\rm D}$ simultaneously.
The correlation between $r$ and $\alpha_{\rm D}$ has disappeared.	

How well was dust cleaned? We have repeated this one-component
foreground cleaning test for various values of $r_{\rm input}$ from 0.001 to
0.1. The results are shown in 
Table~\ref{tab:dustcleaning}: in all cases, the method recovers $r$
successfully. 

\begin{deluxetable}{ccc}
\tablecolumns{3}
\tablecaption{Dust-only Test}
\tablehead{\colhead{$r_{\rm input}$\tablenotemark{a}}&
\colhead{mean($r$)\tablenotemark{b}}&\colhead{std($r$)\tablenotemark{c}}
}
\startdata
0.001 & 0.0011 & 0.0003 \nl
0.003 & 0.0030 & 0.0005 \nl
0.010 & 0.0102 & 0.0010 \nl 
0.030 & 0.0296 & 0.0021 \nl
0.100 & 0.0991 & 0.0057
\enddata
\tablenotetext{a}{Input values of the scalar-to-tensor ratio for
 simulations (64 realizations for each $r_{\rm input}$).}
\tablenotetext{b}{Mean of the recovered maximum likelihood
 values of $r$.} 
\tablenotetext{c}{Standard deviation of the recovered
 maximum likelihood values of $r$.}
\label{tab:dustcleaning}
\end{deluxetable}

\subsection{Cleaning synchrotron in multi-region}
We are now ready to include synchrotron. Our model is
\begin{eqnarray}
\left[Q,U\right](60) &=& \mbox{CMB}+\mbox{Synch(60)}+\mbox{Dust(60)}\\
\nonumber
\left[Q,U\right](100) &=&
\mbox{CMB}+\mbox{Synch(100)}+\mbox{Dust(100)}\\
& &+\mbox{Noise}\\
\left[Q,U\right](240) &=& \mbox{CMB}+\mbox{Synch(240)}+\mbox{Dust(240)}
\end{eqnarray}

It turns out cleaning synchrotron is more challenging than cleaning
dust, as the spatial distribution of synchrotron tends to be more
extended above the Galactic plane than that of dust (see the top panels
of Figure~\ref{fig:foregrounds}).
We start by adding a mock synchrotron model (MSM) map to the PSM dust map.
The MSM map has the same synchrotron polarization intensity across the sky as PSM
at 30~GHz, but has a spatially invariant spectral index of $\beta = -3.0$
(where $\beta=-3.0$ is the average of spatially varying spectral index of PSM). 
With MSM plus PSM dust, $r$ is recoverd successfully;
$\mbox{mean}(r)=0.0012$ and $0.0031$ for $r_{input}=0.001$ and $0.003$.
(see the second and third columns of Table~\ref{tab:synchonly})

Even more problematic is the {\it spatial
variation of the synchrotron spectral index} (see the bottom left panel
of Figure~\ref{fig:foregrounds}), which causes a mismatch
between a template map at 60~GHz and the actual synchrotron distribution
at 100~GHz. 
When we use a single synchrotron
coefficient, $\alpha_{\rm 
S}$, for the whole sky
for the PSM model even without dust ($=$ synchrotron only), we find a 
bias in $r$ of order $\Delta r\approx 0.002$: $\mbox{mean}(r)=0.0028$
and $0.0120$ for  
$r_{\rm input}=0.001$ and 0.01, respectively
(see the fourth and fifth columns of Table~\ref{tab:synchonly}).

\begin{deluxetable*}{ccccccc}
\tablecolumns{7}
\tablewidth{0pt}
\tablecaption{MSM and Synchrotron-only Tests (global and 48~regions)}
\tablehead{\colhead{$r_{\rm input}$\tablenotemark{a}}&
\multicolumn{2}{c}{MSM\tablenotemark{b}} &
\multicolumn{2}{c}{Global\tablenotemark{c}} &
\multicolumn{2}{c}{48~Regions\tablenotemark{c}} \\
\colhead{} &
\colhead{mean($r$)\tablenotemark{d}} &
\colhead{std($r$)\tablenotemark{e}} &
\colhead{mean($r$)\tablenotemark{d}} &
\colhead{std($r$)\tablenotemark{e}} &
\colhead{mean($r$)\tablenotemark{d}} &
\colhead{std($r$)\tablenotemark{e}}}
\startdata
0.001 & 0.0012 & 0.0004 & 0.0028 & 0.0005 & 0.0024 & 0.0005\nl
0.003 & 0.0031 & 0.0006 & 0.0049 & 0.0008 & 0.0046 & 0.0007\nl
0.010 & - & - &           0.0120 & 0.0011 & 0.0115 & 0.0011  
\enddata
\tablenotetext{a}{Input values of the scalar-to-tensor ratio for
 simulations (64 realizations for each $r_{\rm input}$).}
\tablenotetext{b}{MSM plus PSM dust.} 
\tablenotetext{c}{PSM synchrotron only.} 
\tablenotetext{d}{Mean of the recovered maximum likelihood
 values of $r$.} 
\tablenotetext{e}{Standard deviation of the recovered
 maximum likelihood values of $r$.}
\label{tab:synchonly}
\end{deluxetable*}

\begin{figure}[t]
\begin{center}
\includegraphics[width=8cm]{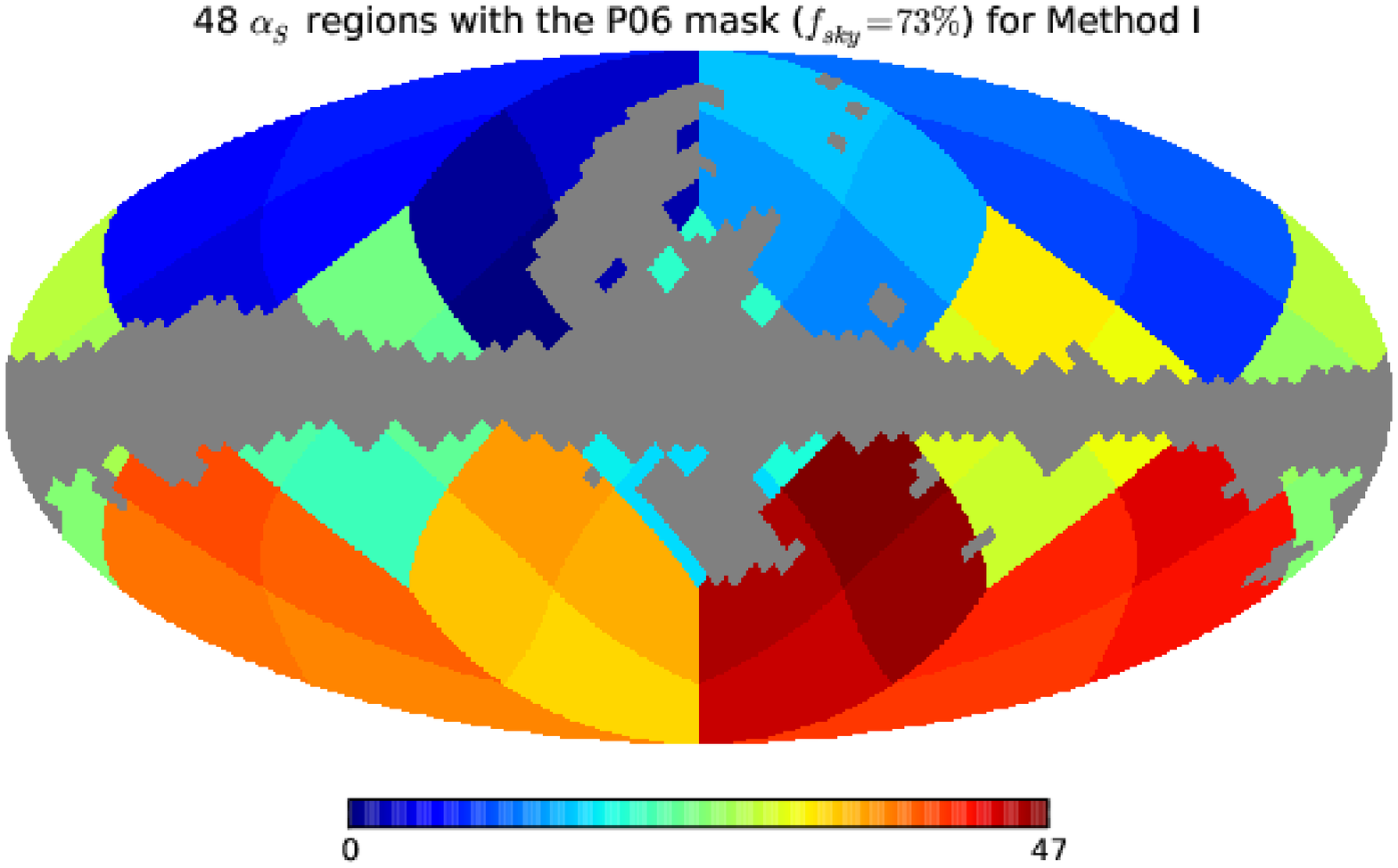}
\includegraphics[width=8cm]{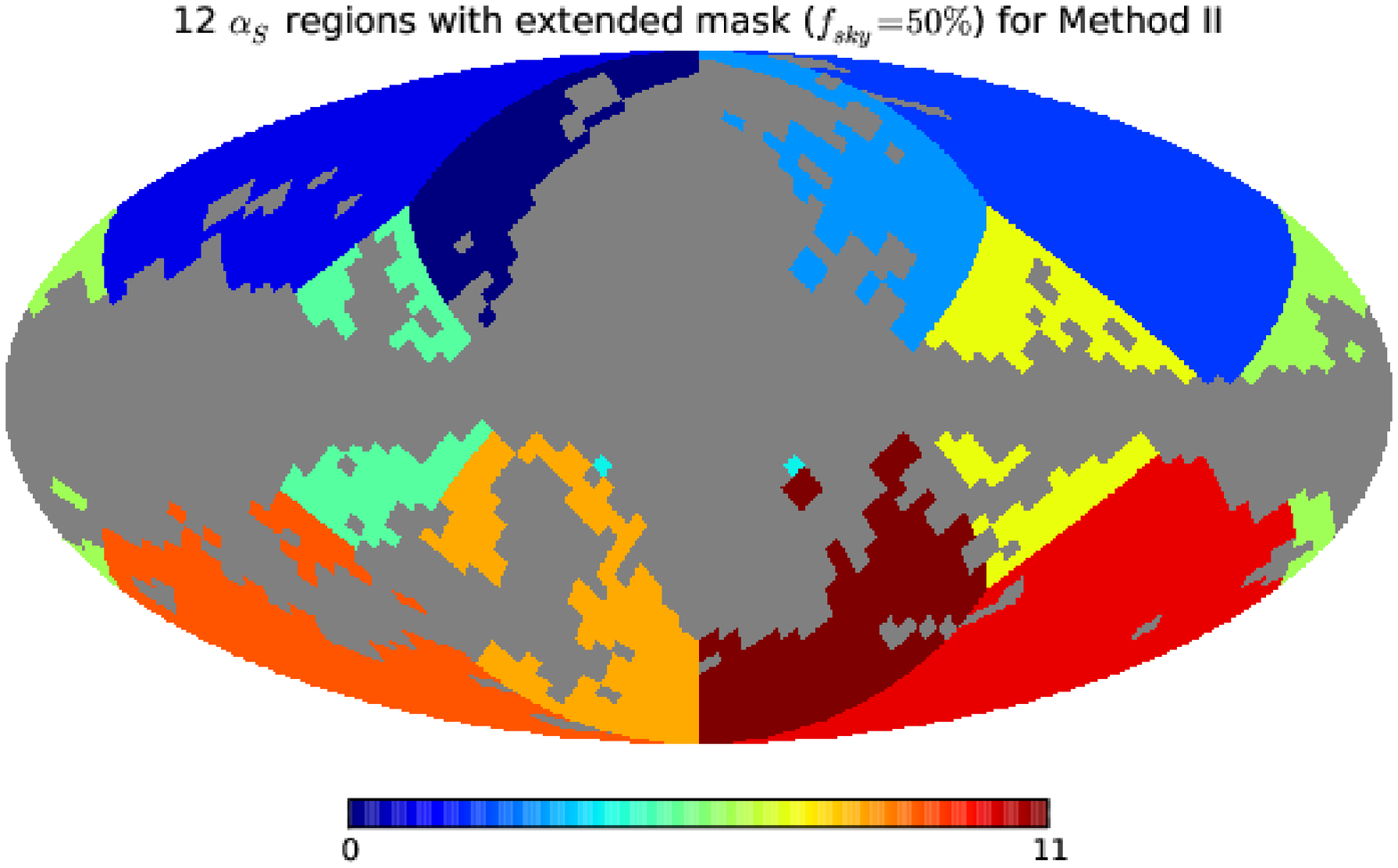}
\caption{
 (a) left:~Method I. The P06 masked sky has been divided into 48 regions based on the Healpix
 map with $N_{\rm side}=2$. $f_{sky}$ of the mask is $73\%$.
 (b) right:~Method II. The extended-masked sky has been divided into 12 regions based on
 the Healpix map with $N_{\rm side}=1$. $f_{sky}$ of the mask is $50\%$.
}
\label{fig:alpha-regions}
\end{center}
\end{figure}

One way to mitigate this issue would be to extend the Galactic mask
\citep{efstathiou/gratton/paci:2009}. In addition, one may give up 
using a single synchrotron amplitude for the whole sky, and use multiple
amplitudes depending on the locations on the
sky.\footnote{Ultimately, 
the best way to mitigate this issue would be to obtain and use information on
the spatial distribution of the synchrotron spectral index.}
In this paper, we
\begin{enumerate}[(Method I)]
\item continue to use the P06 mask, but divide the sky
using the Healpix map with $N_{\rm 
side}=2$, as shown in Figure~\ref{fig:alpha-regions}a and
\item extend the mask to $f_{sky}=50\%$ and divide the sky 
using the Healpix map with $N_{\rm side}=1$, as shown in Figure \ref{fig:alpha-regions}b.
\end{enumerate}
We give the details 
of our definition of the extended mask in Appendix~\ref{sec:mask}.
In short, we choose 
the threshold polarization intensity values at 60 and 240~GHz above which the 
pixels are masked, such that we retain 50\% of the sky.

While we
would probably do a better job at cleaning synchrotron if we divide the
sky according to our knowledge of the polarized synchrotron measured by
{\sl WMAP}, for this paper we prefer to explore a simpler algorithm and
see how well we can recover $r$.

Each region $I$ will be cleaned as (c.f., Equation~(\ref{eq:cleanedmap}))
\begin{equation}
{\bm x}'_I = \frac{[Q_I,U_I](100)-\alpha_{\rm D}[Q_I,U_I](240)-\alpha_{\rm S}^I[Q_I,U_I](60)}{1-\alpha_{\rm D}-\alpha_{\rm S}^I}.
\end{equation}
Note that we still use a single amplitude for dust on the whole
sky. Similarly, the covariance matrix is given by (c.f.,
Equation~(\ref{eq:cleanedcinv})) 
\begin{eqnarray}
\nonumber
{\bm C}_{IJ}(r,s,\alpha_i) &=& r {\bm c}_{IJ}^{\rm tensor} + s {\bm c}_{IJ}^{\rm
 scalar} \\
& &+ \frac{{\bm N}_{1,IJ} + {\bm N}_{2,IJ}}{(1-\alpha_{\rm
 D}-\alpha_{\rm S}^I)(1-\alpha_{\rm
 D}-\alpha_{\rm S}^J)},
\label{eq:covariance}
\end{eqnarray}
where ${\bm C}_{IJ}$, ${\bm c}_{IJ}$  and ${\bm N}_{IJ}$ denote a block
of matrices for 
pixels within regions $I$ and $J$.

The free parameters in the maximization are $r$, $s$, $\alpha_{\rm D}$,
 and $\alpha_{\rm S}^i$ (i=1\ldots 12(48)). In principle we wish to maximize
 the full likelihood function with respect to these parameters; however,
 in practice, this process is too time consuming to do brute-force,
 as varying each of these 15(51) parameters requires re-inverting a $4518\times
 4518$ matrix. Therefore, we make one approximation: we fix
 $\alpha_{\rm 
 D}$ and $\alpha_{\rm S}$ in the covariance matrix
 (Equation~(\ref{eq:covariance})) at the best-fit values, $\alpha_{\rm
 D}^0$ and $\alpha_{\rm S}^{i,0}$. This is a good
 approximation as long as the noise term is sub-dominant compared to the
 dominant scalar $E$-mode signal, which is always the case for our
 low-noise configuration.\footnote{As we have shown in
 Section~\ref{sec:offset}, the foreground amplitudes and the dominant
 scalar $E$ modes are covariant. Therefore, in order to find the
 best-fit $\alpha$s without running the full likelihood, we had to
 ``cheat'' and measure $\alpha$s in maps that do not contain the
 CMB signal or noise. Of course, we cannot do this in real life and thus we will have to come up with an efficient numerical algorithm for maximizing the full
 likelihood without this approximation. We believe that this is doable,
 so this will not be a limiting factor for our method.}
With this approximation,
\begin{equation}
\mathcal{L}(r, s, \alpha_i) \propto \frac{\exp\left[-\frac{1}{2}{\bm x(\alpha_i)}^T{\bm C(r, s; \alpha^0_i)}^{-1}{\bm x(\alpha_i)}\right]}{\sqrt{\vert {\bm C(r, s; \alpha^0_i)}\vert}}, 
\end{equation}
can be maximized with respect to $r$, $s$ and $\alpha_i$ where $i$ runs
from 1 (dust) to 13(49) (synchrotron for 2 to 13(49)). We use the MINUIT package
\citep{james:1998} for the maximization.

In the fourth and fifth columns of
Table~\ref{tab:synchonly}, we show 
the recovered values of $r$ for the {\it synchrotron-only} cases, in
order to see if dividing the sky into 48 regions helps to reduce the
bias in $r$ that we have just seen. We find that the bias has reduced,
but not by much: $\Delta r=0.0018\to 0.0014$, $0.0019\to 0.0016$, and
$0.0020\to 0.0015$ for $r_{\rm input}=0.001$, 0.003, and 0.01, respectively.
This is probably due to the division not being tailored to match the
distribution of synchrotron emission. While we keep this simple division
and do not pursue a more complex division in this paper, we shall come
back to this issue in the future work.

\subsection{Recovering $r$}
\label{sec:recoveringr}
\begin{figure}[t]
\begin{center}
\includegraphics[width=8cm]{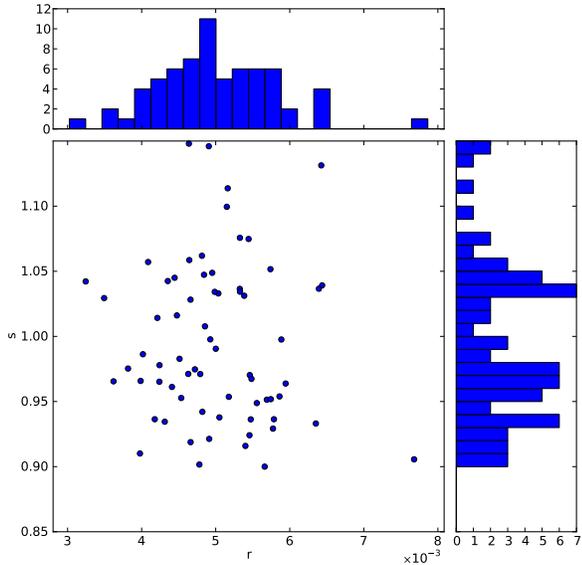}
\caption{%
 Distribution of $r$ and $s$ obtained from 68 realizations $r_{\rm input}=0.003$ using Method I.
}
\label{fig:ds48r0010}
\end{center}
\end{figure}
\begin{deluxetable}{ccccc}
\tablecolumns{5}
\tablecaption{Recovered $r$ from Synchrotron and Dust cleaning}
\tablehead{\colhead{$r_{\rm input}$\tablenotemark{a}}&
\colhead{mean($r$)\tablenotemark{b}}&\colhead{std($r$)\tablenotemark{c}}&
\colhead{mean($r$)\tablenotemark{d}}&\colhead{std($r$)\tablenotemark{e}}
}
\startdata
0.001 & 0.0027 & 0.0005 & 0.0016 & 0.0006\nl
0.003 & 0.0050 & 0.0008 & 0.0038 & 0.0009\nl
0.010 & 0.0121 & 0.0013 & 0.0113 & 0.0015\nl 
0.030 & 0.0326 & 0.0021 & - & - \nl
0.100 & 0.1029 & 0.0053 & - & -
\enddata
\tablenotetext{a}{Input values of the scalar-to-tensor ratio for
 simulations (64(128) realizations for each $r_{\rm input}$ for Method I(II)).}
\tablenotetext{b}{Mean of the recovered maximum likelihood
 values of $r$ for Method I.} 
\tablenotetext{c}{Standard deviation of the recovered
 maximum likelihood values of $r$ for Method I.}
\tablenotetext{d}{Mean of the recovered maximum likelihood
 values of $r$ for Method II.} 
\tablenotetext{e}{Standard deviation of the recovered
 maximum likelihood values of $r$ for Method II.}
\label{tab:fullresult}
\end{deluxetable}
\begin{figure}[t]
\begin{center}
\includegraphics[width=8cm]{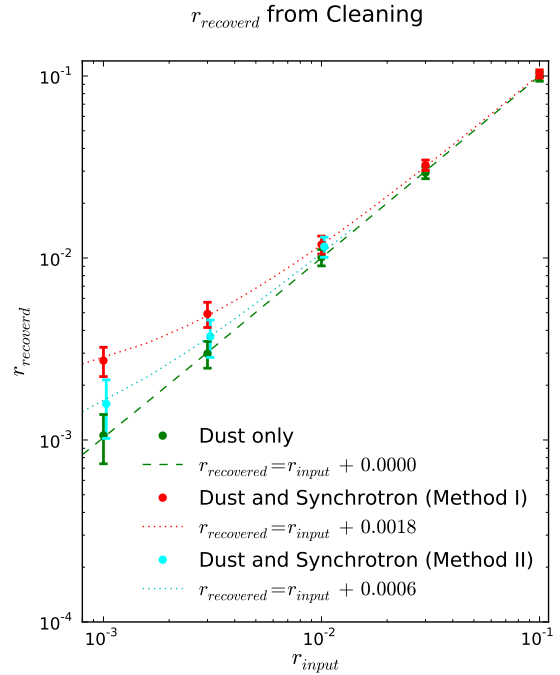}
\caption{%
 Recovered values of $r$ (mean($r$)) and error bars (std($r$)) as a
 function of $r_{\rm input}$. The green points with error bars show the
 recovered $r$ from the dust-only results
 (Table~\ref{tab:dustcleaning}); the red
(cyan)
points with error bars show the
 synchrotron-plus-dust results
using Method I(II)
(Table~\ref{tab:fullresult})
 A systematic bias of $\Delta r\approx 0.002(0.0006)$ is seen for the 
 synchrotron-plus-dust results using Method I(II),
 which can be described by $r_{\rm recovered}=r_{\rm
 input}+(0.0018\pm0.0004)((0.0006\pm0.0004))$ (red(cyan) dotted line). We do not detect an offset for
 the dust-only results: $(0.0000\pm0.0003)$ (green dashed line)
}
\label{fig:r-true-vs-r}
\end{center}
\end{figure}

Now, we recover $r$ from the full dust-plus-synchrotron cases.
In Figure~\ref{fig:ds48r0010}, we show the distribution of $r$ and $s$ for all
of 68 realizations that we have run with $r_{\rm input}=0.003$ using Method I.
In Table~\ref{tab:fullresult}, we show the recovered values of $r$ in
the second and fourth columns. Comparing them to the input values, $r_{\rm input}$,
in the first column, we conclude that our
Method I recovers $r$ with a foreground-induced bias of $\Delta
r\approx 0.002$, 
which is consistent with the bias we have just seen
from the synchrotron-only cases.
With Method II we recover $r$ with a much smaller bias of $\Delta
r\approx 0.0006$.
We visualize our results in Figure~\ref{fig:r-true-vs-r}.

\begin{figure}[t]
\begin{center}
\includegraphics[width=8.5cm]{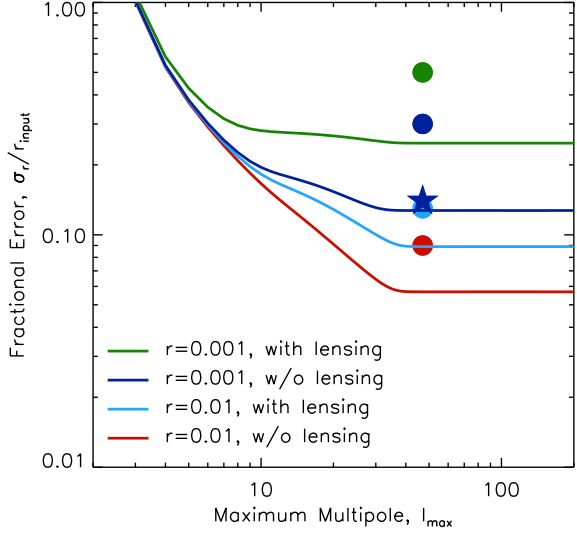}
\caption{%
 The same as Figure~\ref{fig:significance}, but the foreground-free
 predictions are made for simulated maps at $N_{\rm side}=16$ (as
 described in Section~\ref{sec:noise}) and divided by $\sqrt{f_{\rm
 sky}}$ where $f_{\rm sky}=0.735$ for the P06 mask at $N_{\rm
 side}=16$. The filled circles 
 use the error bars, $\sigma_r$, derived from our
 simulations with synchrotron and dust foreground cleaning (denoted as
 ``std($r$)'' in Table~\ref{tab:fullresult}). The denominators, $r_{\rm
 input}$, are not the values recovered from the simulations, but
 the input values fixed at either 0.01 or
 0.001. Each filled circle (foreground-cleaned result from simulations)
 should be 
 compared to the corresponding line with the same color (foreground-free,
 analytical prediction). The star shows the foreground-free and
 lensing-free result 
 from the simulation  with
 $r_{\rm input}=0.001$, which agrees with the analytical prediction to
 within 10\%.
}
\label{fig:significance-with-data}
\end{center}
\end{figure}

The bias in $r$ is important, but the uncertainty in the
recovered $r$ is equally important.
In Figure~\ref{fig:significance}, we have shown 
the predicted fractional errors on the determination of $r$,
$\sigma_r/r_{\rm input}$,
for idealistic full-sky, foreground-free
cases. How would they look when 
synchrotron and dust are included and cleaned with our method? 
In Figure~\ref{fig:significance-with-data}, we show the same figure but
with the predictions made for the simulated CMB-plus-noise maps
at $N_{\rm side}=16$ as described in Section~\ref{sec:noise}, and 
 scaled to the P06 mask. We also show $\sigma_r/r_{\rm input}$ where
 $\sigma_r$ is extracted from the simulations. 
First, when both the
 foreground and lensing noise are ignored, the simulation and the
 analytical prediction are in a good agreement (to within 10\%) for
 $r_{\rm input}=0.001$ (see the star symbol in
 Figure~\ref{fig:significance-with-data}). 
When the foreground is included, however, the error increases.
For $r_{\rm input}=0.01$, 
the foreground cleaning increases the error by about 60\%. 
We see larger discrepancies between the foreground-free
predictions and the foreground-cleaned results for $r_{\rm
input}=0.001$: 
the foreground-cleaned error is a factor of two larger than the
foreground-free prediction; thus, the increase in the error due to  
foreground cleaning can be substantial when $r$ is as small as
$10^{-3}$. 


Further
optimizations could be done, given the details of a
given experiment,
and we intend to explore this issue within the context of some specific
experimental designs. Another improvement can be made by using
$N_{\rm side}=32$ (or 64) map, so that the Nyquist frequency is close to
(or beyond) the second bump of the $B$-mode spectrum and more information
is used. Such analysis, however, would take $2^6=64$ ($4^6=4096$) times
more computation.
\section{Conclusions}
\label{sec:conclusion}
In this paper, we have studied the pixel-based foreground cleaning
method within the context of a next-generation, low-noise CMB polarization
satellite. This method was originally
applied to polarized data
by the {\sl WMAP} team
\citep{page/etal:2007,gold/etal:2009,gold/etal:prep}, and further
investigated by \citet{efstathiou/gratton/paci:2009} in the context of
{\sl Planck}. 

Despite the simplicity of the method (namely, we have maps at 3
different frequencies, two of which are used for removing the
synchrotron and dust emission), we are able to recover the input
tensor-to-scalar ratio with only a small bias, $\Delta r\approx
0.002(<0.001)$ for the P06(extended)
mask, which is dominated by the residual
synchrotron emission.
Further improvements should be straightforward: one can tune the
Galactic mask, and divide the synchrotron fitting regions according to
the actual distribution of the synchrotron spectral index in the Galaxy (rather
than using the regular division shown in
Figure~\ref{fig:alpha-regions}). One may also increase the number of
frequencies for measuring the spatial distribution of the
synchrotron spectral index, provided that we have enough space on the focal
plane. These will be investigated in the 
context of specific experimental designs such as {\sl
LiteBIRD}\footnote{Light satellite for the studies of 
B-mode polarization and Inflation from cosmic background Radiation
Detection; http://cmb.kek.jp/litebird}, and presented elsewhere. 

Our study suggests that a detection of the primordial $B$-mode
polarization at the level of $r\approx 10^{-3}$ should be
possible with carefully optimized mask and $\alpha$ regions. Note that
our statistical error and systematic bias becomes comparable with $f_{sky}=50\%$ mask case.
However,
let us mention one important caveat in our
analysis. While our knowledge of the distribution and properties of the
polarized synchrotron is fairly secure thanks to the {\sl WMAP} data,
our knowledge of the polarized dust emission, especially the
spatial variation of the dust spectral index, is still highly
limited. Therefore, the estimated bias in $r$ that we have presented in
this paper cannot be too accurate. Fortunately, {\sl Planck} will soon
provide us with maps of the polarized dust emission with the
unprecedented sensitivity; thus, we intend to revisit this issue once
the {\sl Planck} data become available. 

\acknowledgements
We thank J. Dunkley for providing us with the Planck Sky Model maps
v1.6.2, and T. Matsumura for providing us with the map of $N_{\rm
obs}$. We acknowledge use of the HEALPix \citep{gorski/etal:2005}, CAMB
\citep{lewis/challinor/lasenby:2000}, and MINUIT \citep{james:1998}
packages. This work was supported by MEXT KAKENHI 21111002 and 22111510.
\appendix
\section{Signal Covariance matrix}
\label{sec:cov}
Given power spectra, $c_\ell^{BB}$ and $c_\ell^{EE}$, components of the
signal covariance 
matrix for $Q$ and $U$ can be computed analytically. We have
$$
{\bm c}(\hat{\bm n},{\hat{\bm n}'})
=\left(
\begin{array}{cc}
c_{QQ}(\hat{\bm n},{\hat{\bm n}'}) & c_{QU}(\hat{\bm n},{\hat{\bm n}'}) \\
c_{UQ}(\hat{\bm n},{\hat{\bm n}'}) & c_{UU}(\hat{\bm n},{\hat{\bm n}'})
\end{array}
\right),
$$
where
\begin{eqnarray*}
c_{QQ}(\hat{\bm n},{\hat{\bm n}'}) & = & \sum_l c_l^{EE}w_l^2 \sum_m W_{lm}(\hat{\bm n})W^*_{lm}({\hat{\bm n}'}) + 
\sum_lc_l^{BB}w_l^2 \sum_m X_{lm}(\hat{\bm n})X^*_{lm}({\hat{\bm n}'}) \\
c_{QU}(\hat{\bm n},{\hat{\bm n}'}) & = & \sum_l c_l^{EE}w_l^2 \sum_m [-W_{lm}(\hat{\bm n})X^*_{lm}({\hat{\bm n}'})] + 
\sum_lc_l^{BB}w_l^2 \sum_m X_{lm}(\hat{\bm n})W^*_{lm}({\hat{\bm n}'}) \\
c_{UQ}(\hat{\bm n},{\hat{\bm n}'}) & = & \sum_l c_l^{EE}w_l^2 \sum_m [-X_{lm}(\hat{\bm n})W^*_{lm}({\hat{\bm n}'})] + 
\sum_lc_l^{BB}w_l^2 \sum_m W_{lm}(\hat{\bm n})X^*_{lm}({\hat{\bm n}'}) \\
c_{UU}(\hat{\bm n},{\hat{\bm n}'}) & = & \sum_l c_l^{EE}w_l^2 \sum_m X_{lm}(\hat{\bm n})X^*_{lm}({\hat{\bm n}'}) + 
\sum_lc_l^{BB}w_l^2 \sum_m W_{lm}(\hat{\bm n})W^*_{lm}({\hat{\bm n}'})
\end{eqnarray*}
and
\begin{eqnarray*}
W_{lm}(\hat{\bm n}) & \equiv & (-1)[_{2}Y_{lm}(\hat{\bm n}) + _{-2}Y_{lm}(\hat{\bm n})]/2, \\
X_{lm}(\hat{\bm n}) & \equiv & (-i)[_{2}Y_{lm}(\hat{\bm n}) - _{-2}Y_{lm}(\hat{\bm n})]/2.
\end{eqnarray*}
We have assumed that $E$ modes and $B$ modes are uncorrelated. 
Here, $w_l$ is a smoothing function which includes an experimental beam,
a pixel window function, and any other smoothing applied to maps.
\section{Extended Mask}
\label{sec:mask}

The resolution 4 ($r4$) mask is extended from the P06 mask by setting
the threshold foreground polarization intensity values at 60 and 240~GHz
above which the  pixels are masked.
The intensity of the pixel $i$ in the resolution 7 map is defined as
\begin{equation}
P_i(\nu)=\sqrt{Q_i^2(\nu)+U_i^2(\nu)}
\end{equation}
where $Q$ and $U$ are the sum of synchrotron and dust:
\begin{equation}
[Q_i, U_i](\nu) = [Q_{i,{\rm synch}}, U_{i,{\rm synch}}](\nu) +  [Q_{i,{\rm dust}}, U_{i,{\rm dust}}](\nu)
\end{equation}
using PSM (See Eqs.~(\ref{eq:synch}) and (\ref{eq:dust})).

An $r4$ pixel is masked if
\begin{enumerate}
\item median of $P_i(240)$ pixels in the $r4$ pixel exceeds Threshold I, or
\item maximum of $P_i(240)$ in the $r4$ pixel exceeds Threshold II, or
\item median of $P_i(60)$ pixels in the $r4$ pixel exceeds Threshold III, or
\item maximum of $P_i(60)$ in the $r4$ pixel exceeds Threshold IV.
\end{enumerate}
Keeping $f_{sky}=50\%$, the values of the four thresholds are determined by
minimizing the total foreground intensity in the residual map;
\begin{equation}
P_{\rm res}({\rm mask}) =
\sum_{i\notin{\rm mask}}\sqrt{Q^2_{{\rm res},i}+U^2_{{\rm res},i}}
\end{equation}
where
\begin{equation}
[Q_{{\rm res}, i},U_{{\rm res}, i}] = [Q_i, U_i](100) - \alpha_D[Q_i, U_i](240) - \alpha_S[Q_i, U_i](60).
\end{equation}
$\alpha_D$ and $\alpha_S$ are given in the usual way by solving
\begin{equation}
\frac{\partial \chi^2}{\partial \alpha_j} = 0, (j = D, S)
\end{equation}
where
\begin{equation}
\chi^2 = [Q_{{\rm res}, i},U_{{\rm res}, i}]^T[Q_{{\rm res}, i}, U_{{\rm res}, i}], (Q_{{\rm res},i} = U_{{\rm res},i} = 0, i \in{\rm mask})
\end{equation}
The median and max.~thresholds for the 240(60)~GHz map determined this way are $19.2(1.42)$ and $38.4(2.11)\mu K.$
I.e., ${\rm Threshold~I} = 19.2$, ${\rm II} = 38.4$, ${\rm III} = 1.42$, and ${\rm IV} = 2.11~\mu$K.  
Note that we have defined an extended mask by using PSM maps without CMB or
noise. In practice, both contributions would add noise spikes to the mask which need to be carefully examined. The noise contribution should be quite small given that we consider a low-noise ($2~\mu$K~arcmin) experiment in this paper. The CMB contribution can be removed by taking the difference between different channels and defining the threshold values on the difference maps (in the same way that the WMAP team has created temperature masks).
However, given $f_{sky}$,
$P_{\rm res}({\rm mask})$ has a very broad bottom as a function of the
thresholds. At the bottom, the shape of the mask is stable and our results are
insensitive to the choice of the threshold values or the algorithm.


\begin{thebibliography}{29}
\expandafter\ifx\csname natexlab\endcsname\relax\def\natexlab#1{#1}\fi

\bibitem[{{Bock} {et~al.}(2008){Bock}, {Cooray}, {Hanany}, {Keating}, {Lee},
  {Matsumura}, {Milligan}, {Ponthieu}, {Renbarger}, \& {Tran}}]{bock/etal:prep}
{Bock}, J., {Cooray}, A., {Hanany}, S., {Keating}, B., {Lee}, A., {Matsumura},
  T., {Milligan}, M., {Ponthieu}, N., {Renbarger}, T., \& {Tran}, H. 2008,
  ArXiv e-prints

\bibitem[{{Brown} {et~al.}(2009){Brown}, {Ade}, {Bock}, {Bowden}, {Cahill},
  {Castro}, {Church}, {Culverhouse}, {Friedman}, {Ganga}, {Gear}, {Gupta},
  {Hinderks}, {Kovac}, {Lange}, {Leitch}, {Melhuish}, {Memari}, {Murphy},
  {Orlando}, {O'Sullivan}, {Piccirillo}, {Pryke}, {Rajguru}, {Rusholme},
  {Schwarz}, {Taylor}, {Thompson}, {Turner}, {Wu}, {Zemcov}, \& {The QUa D
  collaboration}}]{brown/etal:2009}
{Brown}, M.~L., {Ade}, P., {Bock}, J., {Bowden}, M., {Cahill}, G., {Castro},
  P.~G., {Church}, S., {Culverhouse}, T., {Friedman}, R.~B., {Ganga}, K.,
  {Gear}, W.~K., {Gupta}, S., {Hinderks}, J., {Kovac}, J., {Lange}, A.~E.,
  {Leitch}, E., {Melhuish}, S.~J., {Memari}, Y., {Murphy}, J.~A., {Orlando},
  A., {O'Sullivan}, C., {Piccirillo}, L., {Pryke}, C., {Rajguru}, N.,
  {Rusholme}, B., {Schwarz}, R., {Taylor}, A.~N., {Thompson}, K.~L., {Turner},
  A.~H., {Wu}, E.~Y.~S., {Zemcov}, M., \& {The QUa D collaboration}. 2009,
  \apj, 705, 978

\bibitem[{{Chiang} {et~al.}(2010){Chiang}, {Ade}, {Barkats}, {Battle},
  {Bierman}, {Bock}, {Dowell}, {Duband}, {Hivon}, {Holzapfel}, {Hristov},
  {Jones}, {Keating}, {Kovac}, {Kuo}, {Lange}, {Leitch}, {Mason}, {Matsumura},
  {Nguyen}, {Ponthieu}, {Pryke}, {Richter}, {Rocha}, {Sheehy}, {Takahashi},
  {Tolan}, \& {Yoon}}]{chiang/etal:2010}
{Chiang}, H.~C., {Ade}, P.~A.~R., {Barkats}, D., {Battle}, J.~O., {Bierman},
  E.~M., {Bock}, J.~J., {Dowell}, C.~D., {Duband}, L., {Hivon}, E.~F.,
  {Holzapfel}, W.~L., {Hristov}, V.~V., {Jones}, W.~C., {Keating}, B.~G.,
  {Kovac}, J.~M., {Kuo}, C.~L., {Lange}, A.~E., {Leitch}, E.~M., {Mason},
  P.~V., {Matsumura}, T., {Nguyen}, H.~T., {Ponthieu}, N., {Pryke}, C.,
  {Richter}, S., {Rocha}, G., {Sheehy}, C., {Takahashi}, Y.~D., {Tolan}, J.~E.,
  \& {Yoon}, K.~W. 2010, \apj, 711, 1123

\bibitem[{{Chiang} {et~al.}(2008){Chiang}, {Naselsky}, \&
  {Coles}}]{chiang/naselsky/coles:2008}
{Chiang}, L., {Naselsky}, P.~D., \& {Coles}, P. 2008, Modern Physics Letters A,
  23, 1489

\bibitem[{{Coulson} {et~al.}(1994){Coulson}, {Crittenden}, \&
  {Turok}}]{coulson/crittenden/turok:1994}
{Coulson}, D., {Crittenden}, R.~G., \& {Turok}, N.~G. 1994, \prl, 73, 2390

\bibitem[{{Dunkley} {et~al.}(2009){Dunkley}, {Komatsu}, {Nolta}, {Spergel},
  {Larson}, {Hinshaw}, {Page}, {Bennett}, {Gold}, {Jarosik}, {Weiland},
  {Halpern}, {Hill}, {Kogut}, {Limon}, {Meyer}, {Tucker}, {Wollack}, \&
  {Wright}}]{dunkley/etal:2008}
{Dunkley}, J., {Komatsu}, E., {Nolta}, M.~R., {Spergel}, D.~N., {Larson}, D.,
  {Hinshaw}, G., {Page}, L., {Bennett}, C.~L., {Gold}, B., {Jarosik}, N.,
  {Weiland}, J.~L., {Halpern}, M., {Hill}, R.~S., {Kogut}, A., {Limon}, M.,
  {Meyer}, S.~S., {Tucker}, G.~S., {Wollack}, E., \& {Wright}, E.~L. 2009,
  \apjs, 180, 306

\bibitem[{{Efstathiou} {et~al.}(2009){Efstathiou}, {Gratton}, \&
  {Paci}}]{efstathiou/gratton/paci:2009}
{Efstathiou}, G., {Gratton}, S., \& {Paci}, F. 2009, \mnras, 397, 1355

\bibitem[{Finkbeiner {et~al.}(1999)Finkbeiner, Davis, \&
  Schlegel}]{finkbeiner/davis/schlegel:1999}
Finkbeiner, D.~P., Davis, M., \& Schlegel, D.~J. 1999, \apj, 524, 867

\bibitem[{{Fraisse} {et~al.}(2008){Fraisse}, {Brown}, {Dobler}, {Dotson},
  {Draine}, {Frisch}, {Haverkorn}, {Hirata}, {Jansson}, {Lazarian},
  {Magalh{\~a}es}, {Waelkens}, \& {Wolleben}}]{fraisse/etal:prep}
{Fraisse}, A.~A., {Brown}, J., {Dobler}, G., {Dotson}, J.~L., {Draine}, B.~T.,
  {Frisch}, P.~C., {Haverkorn}, M., {Hirata}, C.~M., {Jansson}, R., {Lazarian},
  A., {Magalh{\~a}es}, A.~M., {Waelkens}, A., \& {Wolleben}, M. 2008, ArXiv
  e-prints, arXiv:0811.3920

\bibitem[{{Gold} {et~al.}(2009){Gold}, {Bennett}, {Hill}, {Hinshaw}, {Odegard},
  {Spergel}, {Weiland}, {Dunkley}, {Halpern}, {Jarosik}, {Kogut}, {Komatsu},
  {Larson}, {Meyer}, {Nolta}, {Wollack}, \& {Wright}}]{gold/etal:2009}
{Gold}, B., {Bennett}, C.~L., {Hill}, R.~S., {Hinshaw}, G., {Odegard}, N.,
  {Spergel}, D.~N., {Weiland}, J., {Dunkley}, J., {Halpern}, M., {Jarosik}, N.,
  {Kogut}, A., {Komatsu}, E., {Larson}, D., {Meyer}, S.~S., {Nolta}, M.,
  {Wollack}, E., \& {Wright}, E.~L. 2009, \apjs, 180, 265

\bibitem[{{Gold} {et~al.}(2010)}]{gold/etal:prep}
{Gold}, B. {et~al.} 2010, Astrophys. J. Suppl., submitted

\bibitem[{Gorski {et~al.}(2005)Gorski, Hivon, Banday, Wandelt, Hansen,
  Reinecke, \& Bartlemann}]{gorski/etal:2005}
Gorski, K.~M., Hivon, E., Banday, A.~J., Wandelt, B.~D., Hansen, F.~K.,
  Reinecke, M., \& Bartlemann, M. 2005, \apj, 622, 759

\bibitem[{{Hamimeche} \& {Lewis}(2008)}]{hamimeche/lewis:2008}
{Hamimeche}, S. \& {Lewis}, A. 2008, \prd, 77, 103013

\bibitem[{{Haslam} {et~al.}(1981){Haslam}, {Klein}, {Salter}, {Stoffel},
  {Wilson}, {Cleary}, {Cooke}, \& {Thomasson}}]{haslam/etal:1981}
{Haslam}, C.~G.~T., {Klein}, U., {Salter}, C.~J., {Stoffel}, H., {Wilson},
  W.~E., {Cleary}, M.~N., {Cooke}, D.~J., \& {Thomasson}, P. 1981, \aap, 100,
  209


\bibitem[{{James}(1988)}]{james:1998}
{James}, F. 1988, {MINUIT, Reference Manual, Version 94.1} (CERN, Geneva,
  Switzerland)

\bibitem[{{Kamionkowski} {et~al.}(1997){Kamionkowski}, {Kosowsky}, \&
  {Stebbins}}]{kamionkowski/kosowsky/stebbins:1997}
{Kamionkowski}, M., {Kosowsky}, A., \& {Stebbins}, A. 1997, \prd, 55, 7368

\bibitem[{{Komatsu} {et~al.}(2010)}]{komatsu/etal:prep}
{Komatsu}, E. {et~al.} 2010, Astrophys. J. Suppl., submitted, arXiv:1001.4538

\bibitem[{{Larson} {et~al.}(2010)}]{larson/etal:prep}
{Larson}, D. {et~al.} 2010, Astrophys. J. Suppl., submitted

\bibitem[{{Leach} {et~al.}(2008){Leach}, {Cardoso}, {Baccigalupi}, {Barreiro},
  {Betoule}, {Bobin}, {Bonaldi}, {Delabrouille}, {de Zotti}, {Dickinson},
  {Eriksen}, {Gonz{\'a}lez-Nuevo}, {Hansen}, {Herranz}, {Le Jeune},
  {L{\'o}pez-Caniego}, {Mart{\'{\i}}nez-Gonz{\'a}lez}, {Massardi}, {Melin},
  {Miville-Desch{\^e}nes}, {Patanchon}, {Prunet}, {Ricciardi}, {Salerno},
  {Sanz}, {Starck}, {Stivoli}, {Stolyarov}, {Stompor}, \&
  {Vielva}}]{leach/etal:2008}
{Leach}, S.~M., {Cardoso}, J., {Baccigalupi}, C., {Barreiro}, R.~B., {Betoule},
  M., {Bobin}, J., {Bonaldi}, A., {Delabrouille}, J., {de Zotti}, G.,
  {Dickinson}, C., {Eriksen}, H.~K., {Gonz{\'a}lez-Nuevo}, J., {Hansen}, F.~K.,
  {Herranz}, D., {Le Jeune}, M., {L{\'o}pez-Caniego}, M.,
  {Mart{\'{\i}}nez-Gonz{\'a}lez}, E., {Massardi}, M., {Melin}, J.,
  {Miville-Desch{\^e}nes}, M., {Patanchon}, G., {Prunet}, S., {Ricciardi}, S.,
  {Salerno}, E., {Sanz}, J.~L., {Starck}, J., {Stivoli}, F., {Stolyarov}, V.,
  {Stompor}, R., \& {Vielva}, P. 2008, \aap, 491, 597

\bibitem[{{Lewis} {et~al.}(2000){Lewis}, {Challinor}, \&
  {Lasenby}}]{lewis/challinor/lasenby:2000}
{Lewis}, A., {Challinor}, A., \& {Lasenby}, A. 2000, \apj, 538, 473

\bibitem[{Liddle \& Lyth(2009)}]{liddle/lyth:PDP}
Liddle, A.~R. \& Lyth, D.~H. 2009, The Primordial Density Perturbation:
  Cosmology, Inflation and the Origin of Structure (Cambridge University Press)

\bibitem[{{Miville-Desch{\^e}nes} {et~al.}(2008){Miville-Desch{\^e}nes},
  {Ysard}, {Lavabre}, {Ponthieu}, {Mac{\'{\i}}as-P{\'e}rez}, {Aumont}, \&
  {Bernard}}]{miville-deschenes/etal:2008}
{Miville-Desch{\^e}nes}, M., {Ysard}, N., {Lavabre}, A., {Ponthieu}, N.,
  {Mac{\'{\i}}as-P{\'e}rez}, J.~F., {Aumont}, J., \& {Bernard}, J.~P. 2008,
  \aap, 490, 1093

\bibitem[{{Page} {et~al.}(2007){Page}, {Hinshaw}, {Komatsu}, {Nolta},
  {Spergel}, {Bennett}, {Barnes}, {Bean}, {Dor{\'e}}, {Dunkley}, {Halpern},
  {Hill}, {Jarosik}, {Kogut}, {Limon}, {Meyer}, {Odegard}, {Peiris}, {Tucker},
  {Verde}, {Weiland}, {Wollack}, \& {Wright}}]{page/etal:2007}
{Page}, L., {Hinshaw}, G., {Komatsu}, E., {Nolta}, M.~R., {Spergel}, D.~N.,
  {Bennett}, C.~L., {Barnes}, C., {Bean}, R., {Dor{\'e}}, O., {Dunkley}, J.,
  {Halpern}, M., {Hill}, R.~S., {Jarosik}, N., {Kogut}, A., {Limon}, M.,
  {Meyer}, S.~S., {Odegard}, N., {Peiris}, H.~V., {Tucker}, G.~S., {Verde}, L.,
  {Weiland}, J.~L., {Wollack}, E., \& {Wright}, E.~L. 2007, \apjs, 170, 335


\bibitem[{Planck Blue Book} (2005)]{planck:bb}
{Plank; The Scientific Program} ESA-SCI(2005)1

\bibitem[{QUIET} (2010)]{quiet:prep}
{QUIET} Collaboration: {Bischoff}, C., {Brizius}, A., {Buder}, I., {Chinone}, Y., {Cleary},
  K., {Dumoulin}, R.~N., {Kusaka}, A., {Monsalve}, R., {N{\ae}ss}, S.~K.,
  {Newburgh}, L.~B., {Reeves}, R., {Smith}, K.~M., {Wehus}, I.~K., {Zuntz},
  J.~A., {Zwart}, J.~T.~L., {Bronfman}, L., {Bustos}, R., {Church}, S.~E.,
  {Dickinson}, C., {Eriksen}, H.~K., {Ferreira}, P.~G., {Gaier}, T.,
  {Gundersen}, J.~O., {Hasegawa}, M., {Hazumi}, M., {Huffenberger}, K.~M.,
  {Jones}, M.~E., {Kangaslahti}, P., {Kapner}, D.~J., {Lawrence}, C.~R.,
  {Limon}, M., {May}, J., {McMahon}, J.~J., {Miller}, A.~D., {Nguyen}, H.,
  {Nixon}, G.~W., {Pearson}, T.~J., {Piccirillo}, L., {Radford}, S.~J.~E.,
  {Readhead}, A.~C.~S., {Richards}, J.~L., {Samtleben}, D., {Seiffert}, M.,
  {Shepherd}, M.~C., {Staggs}, S.~T., {Tajima}, O., {Thompson}, K.~L.,
  {Vanderlinde}, K., {Williamson}, R., \& {Winstein}, B. 2010, ArXiv e-prints,
  arXiv:1012.3191

\bibitem[{{Seljak} \& {Zaldarriaga}(1996)}]{seljak/zaldarriaga:1996}
{Seljak}, U. \& {Zaldarriaga}, M. 1996, \apj, 469, 437

\bibitem[{Seljak \& Zaldarriaga(1997)}]{seljak/zaldarriaga:1997}
Seljak, U. \& Zaldarriaga, M. 1997, Phys. Rev. Lett., 78, 2054

\bibitem[{Weinberg(2008)}]{weinberg:COS}
Weinberg, S. 2008, Cosmology (Oxford, UK: Oxford University Press)

\bibitem[{{Zaldarriaga} {et~al.}(2008){Zaldarriaga}, {Colombo}, {Komatsu},
  {Lidz}, {Mortonson}, {Oh}, {Pierpaoli}, {Verde}, \&
  {Zahn}}]{zaldarriaga/etal:prep}
{Zaldarriaga}, M., {Colombo}, L., {Komatsu}, E., {Lidz}, A., {Mortonson}, M.,
  {Oh}, S.~P., {Pierpaoli}, E., {Verde}, L., \& {Zahn}, O. 2008, ArXiv
  e-prints, arXiv:0811.3918

\bibitem[{Zaldarriaga \& Seljak(1998)}]{zaldarriaga/seljak:1998}
Zaldarriaga, M. \& Seljak, U. 1998, Phys. Rev., D58, 023003

\end{thebibliography}

\end{document}